\documentclass{article}
\usepackage{curves}
\usepackage{latexsym,amssymb,graphicx}

\newcommand{\fract}[2]{(#1/#2)} %fraction in text
\newcommand{\leftn}{} %don't use big brackets, to save space
\newcommand{\rightn}{} %don't use big brackets, to save space

\newcommand{\R}{{\mathbb R}}  %ams bold
\newcommand{\tanCpK}{{\mathcal T}_pK}

\newcommand{\ip}[2]{\langle #1 , #2 \rangle}
\newcommand{\co}{\mbox{co}\,}
\newcommand{\Xo}{{\mathcal V}}
\newcommand{\So}{{\mathcal S}}
\newcommand{\tanSx}{{\mathcal T}_\xi \So}
\newcommand{\convS}{\raisebox{-1.5ex}{$\,\stackrel{\rightarrow }{{\scriptstyle\So}}\,$}}
\newcommand{\Sop}{{\mathcal S}'}
\newcommand{\tanSpx}{{\mathcal T}_\xi \Sop}

\newcommand{\hatF}{\widehat{F}}
\newcommand{\inter}{\mbox{int}\,}
\newcommand{\clos}{\mbox{clos}\,}
\newcommand{\Inpt}{\Inputset^{[2]}}
\newcommand{\Funcs}[1]{#1_\infty }
\newcommand{\Inpf}{\Funcs\Inputset}
\newcommand{\Inpft}{\Funcs\Inpt}

\newcommand{\ybar}{\bar{y}}
\newcommand{\zbar}{\bar{z}}
\newcommand{\ubar}{\bar{u}}
\newcommand{\vbar}{\bar{v}}
\newcommand{\xbar}{\bar{x}}
\newcommand{\ft}{f^{[2]}}
\newcommand{\fit}{\phi ^{[2]}}
\newcommand{\tanxdiffK}{{\cal T}_{\xi _1-\xi _2}K}
\newcommand{\tanxGo}{{\cal T}_{\xi }\Gamma _0}
\newcommand{\tanxG}{{\cal T}_{\xi }\Gamma }
\newcommand{\Uo}{{\mathcal W}}

\def\ben{\begin{enumerate}}
\def\een{\end{enumerate}}

\newtheorem{theorem}{Theorem}
\newtheorem{itlemma}{Lemma}[section] %number by section (set in \em by default)
\newtheorem{itproposition}[itlemma]{Proposition}
\newtheorem{itcorollary}[itlemma]{Corollary}
\newtheorem{itremark}[itlemma]{Remark}
\newtheorem{itdefinition}[itlemma]{Definition}
\newtheorem{itexample}[itlemma]{Example}

\newenvironment{lemma}{\begin{itlemma}\rm}{\end{itlemma}} %no-italics
\newenvironment{remark}{\begin{itremark}\rm}{\end{itremark}} %no-italics
\newenvironment{corollary}{\begin{itcorollary}\rm}{\end{itcorollary}}
\newenvironment{proposition}{\begin{itproposition}\rm}{\end{itproposition}}
\newenvironment{definition}{\begin{itdefinition}\rm}{\end{itdefinition}}
\newenvironment{example}{\begin{itexample}\rm}{\end{itexample}}

\newcommand{\text}[1]{\hbox{\rm \ #1\ \/}}
\newcommand{\be}[1]{\begin{equation}\label{#1}}
\newcommand{\ee}{\end{equation}}
\newcommand{\beqn}{\begin{eqnarray*}}
\newcommand{\eeqn}{\end{eqnarray*}}
\newcommand{\beq}{\begin{eqnarray}}
\newcommand{\eeq}{\end{eqnarray}}
\newcommand{\bl}[1]{\begin{lemma}\label{#1}}
\newcommand{\ble}[1]{\begin{lemmaex}\label{#1}}
\newcommand{\br}[1]{\begin{remark}\label{#1}}
\newcommand{\bt}[1]{\begin{theorem}\label{#1}}
\newcommand{\bd}[1]{\begin{definition}\label{#1}}
\newcommand{\bp}[1]{\begin{proposition}\label{#1}}
\newcommand{\bc}[1]{\begin{corollary}\label{#1}}
\newcommand{\bfact}[1]{\begin{fact}\label{#1}}
\newcommand{\ber}[1]{\begin{exercise}\label{#1}}
\newcommand{\bex}[1]{\begin{example}\label{#1}}
\newcommand{\bem}[1]{\begin{example}\label{#1}}  %Yes, 2 different ones...
\newcommand{\ec}{\mybox\end{corollary}}
\newcommand{\efact}{\mybox\end{fact}}
\newcommand{\eer}{\mybox\end{exercise}}
\newcommand{\eex}{\mybox\end{example}}
\newcommand{\eem}{\mybox\end{example}}
\newcommand{\el}{\mybox\end{lemma}}
\newcommand{\ele}{\mybox\end{lemmaex}}
\newcommand{\er}{\mybox\end{remark}}
\newcommand{\et}{\qed\end{theorem}}
\newcommand{\ed}{\mybox\end{definition}}
\newcommand{\ep}{\mybox\end{proposition}}
\newcommand{\epr}{\end{proof}}
\newcommand{\bpr}{\begin{proof}}

\newcommand{\ecs}{\end{corollary}}
\newcommand{\eers}{\end{exercise}}
\newcommand{\eexs}{\end{example}}
\newcommand{\eems}{\end{example}}
\newcommand{\els}{\end{lemma}}
\newcommand{\eles}{\end{lemmaex}}
\newcommand{\ers}{\end{remark}}
\newcommand{\ets}{\end{theorem}}
\newcommand{\eds}{\end{definition}}
\newcommand{\eps}{\end{proposition}}
\newcommand{\halmos}{\rule{1ex}{1.4ex}}
\newcommand{\qed}{\hfill \halmos} %put \qed at right margin
\newcommand{\mybox}{\hfill $\Box$} %put \qed at right margin (white square)

\newcommand{\abs}[1]{\left\vert #1 \right\vert}
\newcommand{\ki}{\mathcal{K}_\infty }

\newcommand{\banach}{{\mathbb B}}
\newcommand{\st}{\; | \;}
\newcommand{\Bu}{\banach_\Inputset}
\newcommand{\By}{\banach_\Outputset}
\newcommand{\Inputset}{{\mathcal U}}
\newcommand{\Outputset}{{\mathcal Y}}
\newcommand{\edeq}{\vskip-0.7cm\mybox\end{definition}}

\def\bi{\begin{itemize}}
\def\ei{\end{itemize}}
\newcommand{\coefV}{v}
\newcommand{\coefG}{g}
\newcommand{\coefK}{k}
\newcommand{\coefk}{\kappa }

\newcommand{\differential}{{\mbox{D}}}

\begin{document}

\title{Monotone Control Systems}

\author{David Angeli\thanks{Email: {\tt angeli@dsi.unifi.it}}\\
Dip. Sistemi e Informatica,
University of Florence, 50139 Firenze, Italy\\
Eduardo D. Sontag\thanks{Supported in part by US Air
Force Grant F49620-01-1-0063
and by NIH Grant R01 GM46383.
Email: {\tt sontag@hilbert.rutgers.edu}} \\
Dept. of Mathematics, Rutgers University, NJ 08854, USA}

\date{}
\maketitle

\begin{abstract}
Monotone systems constitute one of the most important classes of
dynamical systems used in mathematical biology modeling.
The objective of this paper is to extend the notion of monotonicity to
systems with inputs
and outputs, a necessary first step in trying to understand
interconnections, especially including feedback loops, built up out of
monotone components. 
Basic definitions and theorems are provided, as well as an application to the
study of a model of one of the cell's most important subsystems.
\end{abstract}

\section{Introduction}

One of the most important classes of dynamical systems in theoretical biology
is that of {\em monotone systems\/}.  Among the classical references in this
area are the textbook by Smith~\cite{Smith} and the papers~\cite{Hirsh,Hirsh2}
by Hirsh and \cite{smale} by Smale.  Monotone systems are those for which
trajectories preserve a partial ordering on states.  
They include the subclass of {\em cooperative\/}
systems (see e.g.~\cite{deleenheer1,deleenheer2,deleenheer3} for recent
contributions in the control literature), for which different state variables
reinforce each other (positive feedback)
as well as more general systems in which each pair of variables may affect
each other in either positive or negative,
or even mixed, forms 
(precise definitions are given below).
Although one may consider systems in which constant parameters (which
can be thought of as constant inputs) appear, as done in the recent
paper \cite{piccardi-rinaldi} for cooperative systems, the concept of
monotone system has been traditionally defined only for systems with
no external input (or ``control'') {\em functions}.

The objective of this paper is to extend the notion of monotone systems to
{\em systems with inputs and outputs\/}. 
This is by no means a purely academic exercise, but it is a necessary 
first step in trying to understand interconnections,
especially including feedback loops, built up out of monotone components.

The successes of systems theory have been due in large part to its ability to
analyze complicated structures on the basis of the behavior of elementary
subsystems, each of which is ``nice'' in a suitable input/output sense
(stable, passive, etc), in conjunction with the use of tools such as the small
gain theorem to characterize interconnections.

On the other hand, one of the main themes and challenges in current molecular
biology lies in the understanding of cell behavior in terms of cascade and
feedback interconnections of elementary ``modules'' which appear
repeatedly, see e.g.~\cite{modules}.  Our work reported here was motivated by
the problem of studying one such module type (closely related to, but more
general than, the example which motivated~\cite{cauchy}), and the realization
that the theory of monotone systems, when extended to allow for inputs,
provides an 
appropriate
tool to formulate and prove basic properties of such
modules.

The organization of this paper is as follows.
In Section~\ref{Monotone Systems}, we introduce the basic concepts,
including the special case of cooperative systems.
Section~\ref{Infinitesimal Characterizations} provides infinitesimal
characterizations of monotonicity, relying upon certain technical points
discussed in the Appendix.
Cascades are the focus of Section~\ref{Cascades of monotone systems},
and Section~\ref{Static Input/State and Input/Output Characteristics}
introduces the notions of 
static Input/State and Input/Output characteristics, which then play a central
role in the study of feedback interconnections and a small-gain theorem
--- the main result in the paper ---
in Section~\ref{Feedback Interconnections of Monotone Systems}.
We return to the biological example of MAPK cascades in
Section~\ref{An Application}.
Finally, Section~\ref{Relations to Positivity} shows the equivalence between
cooperative systems and positivity of linearizations.

We view this paper as only the beginning of a what should be a fruitful
direction of research into a new type of nonlinear systems.  In particular,
in~\cite{angeli-ferrell-sontag} and~\cite{part2}, we present results dealing
with positive feedback interconnections and multiple steady states, and
associated hysteresis behavior, as well as graphical criteria for monotonicity,
and in~\cite{deleenheer4,deleenheer5} we describe applications to population
dynamics and to the analysis of chemostats.

\section{Monotone Systems}
\label{Monotone Systems}

Monotone dynamical systems are usually defined on subsets of ordered
Banach
(or even more general metric) spaces.
An {\em ordered Banach space} is a real Banach space $\banach$ 
together with a distinguished nonempty closed subset $K$ of $\banach$, its
{\em positive cone\/}.
(The spaces $\banach$ which we study in this paper will all be Euclidean
spaces; however, the basic definitions can be given in more generality, and
doing so might eventually be useful for applications such as the study of
systems with delays, as done in~\cite{Smith} for systems without inputs.) 
The set $K$ is assumed to have the following properties: it is a cone, i.e.\ 
$\alpha K \subset K$ for $\alpha \in \R_+$, it is convex (equivalently, since $K$ is a
cone, $K+K \subset K$), 
and pointed, i.e.\ $K \cap (-K)= \{ 0 \}$.
An ordering is then defined by 
$x_1 \succeq x_2$ iff $x_1-x_2 \in K$.
Strict ordering is denoted by $x_1 \succ x_2$, meaning that 
$x_1 \succeq x_2$ and $x_1 \neq x_2$.
One often uses as well the notations $\prec$ and $\preceq$, in the obvious
sense ($x_2 \preceq x_1$ means $x_1 \succeq x_2$).
(Most of the results discussed in this paper use only that $K$ is a cone.  The
property $K \cap (-K)= \{ 0 \}$, which translates into reflexivity of the
order, is used only at one point, and the convexity property, which translates into
transitivity of the order, will be only used in a few places.)

The most typical example would be $\banach=\R^n$ and $K = \R_{\geq 0}^n$, in
which case ``$x_1 \succeq x_2$'' means that each coordinate of $x_1$ is bigger
or equal than the corresponding coordinate of $x_2$.  This order on state
spaces gives rise to the class of ``cooperative systems'' discussed below.
However, other orthants in $\R^n$ other than the positive orthant $K =
\R_{\geq 0}^n$  are often more natural in applications, as we will see.

In view of our interest in biological and chemical applications, we must
allow state spaces to be non-linear subsets of linear spaces.  
For example,
state variables typically represent concentrations, and hence must be
positive, and are often subject to additional inequality constraints such as
stoichiometry or mass preservation.
Thus, from now on, we will assume given an ordered Banach space $\banach$ and a
subset $X$ of $\banach$ which is the closure of an open subset of $\banach$.
For instance, $X=\banach$, or, in an example to be considered later,
$\banach=\R^2$ with the order induced by $K=\R_{\geq 0} \times \R_{\leq 0}$,
and $X=\{(x,y)\in \R^2\st x\geq 0, y\geq 0, x+y\leq 1\}$.  

The standard concept of monotonicity for uncontrolled
systems
is as follows:
A dynamical system $\phi : \R_{\geq 0} \times X \rightarrow X$ is {\em monotone\/}
if this implication holds:
$ x_1 \succeq x_2 \, \Rightarrow \, \phi (t,x_1) \succeq \phi (t,x_2)$
for all $t \geq 0$.
If the positive cone $K$ is {\em solid}, i.e.\ it has a nonempty interior
(as is often the case in applications of monotonicity, see e.g.~\cite{part2})
one can also define 
a stricter ordering: $x_1 \gg x_2 \Leftrightarrow x-y \in \textrm{int}(K)$. 
(For example, when $K = \R_{\geq 0}^n$, this means that every coordinate of
$x_1$ is strictly larger than the corresponding coordinate of $x_2$, in
contrast to ``$x_1 \succ x_2$'' which means merely that some coordinate
is strictly bigger while the rest are bigger or equal.)
Accordingly, one says that a
dynamical system $\phi : \R_{\geq 0} \times X \rightarrow X$ is {\em strongly monotone\/}
if $x_1 \succ x_2$
implies that
$\phi (t,x_1) \gg \phi (t,x_2)$
for all $t\geq 0$.

Next we generalize, in a very natural way, the above definition to 
\emph{controlled} dynamical systems, i.e., systems forced by some exogenous
input signal.    
In order to do so, we assume given a partially ordered input value space
$\Inputset$.
Technically, we will assume that $\Inputset$ is a 
subset of an ordered Banach space $\Bu$.
Thus, for any pair of input values $u_1$ and $u_2 \in \Inputset$,
we write $u_1 \succeq u_2$ whenever $u_1 - u_2 \in K_u$ where $K_u$ is the
corresponding positivity cone in $\Bu$.
In order to keep the notations simple, here and later,
when there is no
risk of ambiguity, we use the same symbol ($\succeq$) to denote ordered pairs
of input values or pairs of states. 

By an ``input'' or ``control'' we shall mean a Lebesgue measurable function
$u(\cdot ):\R_{\geq 0}\rightarrow \Inputset$ which is essentially bounded, i.e.\ 
there is for each finite interval $[0,T]$ some compact subset $C\subseteq \Inputset$
such that $u(t)\in  C$ for almost all $t\in [0,T]$.
We denote by $\Inpf$ the set of all inputs.
Accordingly, given two 
$u_1, u_2\in \Inpf$,
we write
$u_1  \succeq u_2$
if
$u_1 (t) \succeq u_2(t)$
for all $t \geq 0$.
(To be more precise, this and other definitions should be interpreted
in an ``almost everywhere'' sense, since inputs are Lebesgue-measurable
functions.)
A {\em controlled dynamical system} is specified by a state space $X$ as
above, an input set $\Inputset$, and a mapping 
$\phi : \R_{\geq 0} \times X \times \Inpf \rightarrow X$
such that the usual semigroup properties hold.
(Namely, $\phi (0,x,u)=x$ and $\phi (t,\phi (s,x,u_1),u_2)=\phi (s+t,x,v)$, where
$v$ is the restriction of $u_1$ to the interval $[0,s]$ concatenated with
$u_2$ shifted to $[s,\infty )$; we will soon specialize to solutions of
controlled differential equations.)

We interpret $\phi (t,\xi ,u)$ as the state at time $t$ obtained if the initial
state is $\xi $ and the external input is $u(\cdot )$.
Sometimes, when clear from the context, we write ``$x(t,\xi ,u)$'' or just
``$x(t)$'' instead of $\phi (t,\xi ,u)$.
When there is no risk of confusion, we use ``$x$'' to denote states (i.e.,
elements of $X$) as well as trajectories, but for emphasis we sometimes use
$\xi $, possibly subscripted, and other Greek letters, to denote states.
Similarly, ``$u$'' may refer to an input value (element of $\Inputset$)
or an input function (element of $\Inpf$).

\bd{contrmondynsys}
A controlled dynamical system $\phi : \R_{\geq 0} \times X \times 
\Inpf \rightarrow X$ is {\em monotone}
if the implication below holds for all $t \geq 0$:
\[
u_1 \succeq u_2, \; x_1 \succeq x_2 \quad \Rightarrow  \quad 
\phi (t,x_1,u_1) \succeq \phi (t,x_2,u_2).
\]
\edeq

Viewing systems with no inputs as controlled systems for which the input value
space $\Inputset$ has just one element, one recovers the classical definition.
This allows application of the rich theory developed for this class of systems,
such as theorems guaranteeing convergence to equilibria of almost all
trajectories, for strongly monotone systems (defined in complete analogy to the
concept for systems with no inputs);
see~\cite{angeli-ferrell-sontag,part2}.

We will also consider monotone systems {\em with outputs} $y=h(x)$.
These are specified by a controlled monotone system $\phi $ together with a
monotone ($x_1\succeq x_2$ $\Rightarrow $ $h(x_1)\succeq h(x_2)$)
map $h:X\rightarrow \Outputset$, where $\Outputset$, the set of measurement or output
values, is a subset of some ordered Banach space $\By$.
We often use the shorthand $y(t,x,u)$ instead of $h(\phi (t,x,u))$, to denote the
output at time $t$ corresponding to the state obtained from initial state $x$
and input $u$.

From now on, we will specialize to the case of systems defined by differential
equations with inputs:
\be{cds}
\dot x \,=\, f(x,u)
\ee
(see~\cite{mct} for basic definitions and properties regarding such systems).
We make the following technical assumptions.
The map $f$ is defined on $\widetilde X\times \Inputset$, 
where $\widetilde X$ is some open subset of
$\banach$ which contains $X$, and $\banach=\R^n$ for some integer $n$.
We assume that $f(x,u)$ is continuous in $(x,u)$ and locally Lipschitz
continuous in $x$ locally uniformly on $u$.  This last property means that for
each compact subsets $C_1\subseteq  X$ and $C_2\subseteq \Inputset$ there exists some constant
$k$ such that $\abs{f(\xi ,u)-f(\zeta ,u)}\leq k \abs{\xi -\zeta }$ for all $\xi ,\zeta \in  C_1$ and
all $u\in  C_2$. 
(When studying interconections, we will also implicitly assume that $f$ is
locally Lipschitz in $(x,u)$, so that the full system has unique solutions.)
In order to obtain a well-defined controlled dynamical system on $X$,
we will assume that the solution $x(t)=\phi (t,x_0,u)$ of $\dot x=f(x,u)$ with
initial condition $x(0)=x_0$ is defined for all inputs $u(\cdot )$ and all times
$t\geq 0$.
This means that solutions with initial states in $X$ must be defined for all
$t\geq 0$ (forward completeness) and that the set $X$ is forward invariant.
(Forward invariance of $X$ may be checked using tangent cones at the boundary
of $X$, see the Appendix.)

{\em From now on, all systems will be assumed to be of this form.}

\section{Infinitesimal Characterizations}
\label{Infinitesimal Characterizations}

For systems~(\ref{cds}) defined by controlled differential equations,
we will provide an infinitesimal characterization of monotonicity, expressed
directly in terms of the vector field, which does not require the explicit
computation of solutions.
Our result will generalize the well-known Kamke conditions, discussed
in~\cite{Smith}, Chapter 3.
We denote $\Xo:=\inter X$, the interior of $X$ (recall that $X$ is the
closure of $\Xo$) and impose the following 
{\em approximability property} (see~\cite{Smith}, Remark 3.1.4):
{\em for all $\xi _1,\xi _2\in  X$ such that $\xi _1\succeq\xi _2$,
there exist sequences $\{\xi ^i_1\},\{\xi ^i_2\}\subseteq \Xo$
such that
$\xi ^i_1\succeq\xi ^i_2$ for all $i$
and $\xi ^i_1\rightarrow \xi _1$ and $\xi ^i_2\rightarrow \xi _2$ as $i\rightarrow \infty $.}

\br{convex-OK}
The approximability assumption is very mild.  It is satisfied, in particular,
if the set $X$ is convex, and, even more generally, if it is strictly
star-shaped with respect to some interior point $\xi _*$, i.e., for all
$\xi \in  X$ and all $0\leq \lambda <1$, it holds that $\lambda \xi +(1-\lambda )\xi _*\in  \Xo$.
(Convex sets with nonempty interior have this property with respect to any
point $\xi _*\in \Xo$, since 
$\lambda \xi +(1-\lambda )\xi _*\in Q:=\lambda \xi +(1-\lambda )\Xo\subseteq  X$ (the inclusion by convexity) and
the set $Q$ is open because $\eta \mapsto \lambda \xi +(1-\lambda )\eta $ is an invertible affine
mapping.) 
Indeed, suppose that $\xi _1-\xi _2\in  K$, pick any sequence
$\lambda ^i\nearrow1$, and define $\xi ^i_j:=\lambda ^i\xi _j+(1-\lambda ^i)\xi _*$ for $j=1,2$.
These elements are in $\Xo$, they converge to $\xi _1$ and $\xi _2$ respectively,
and each $\xi ^i_1-\xi ^i_2=\lambda ^i(\xi _1-\xi _2)$ belongs to $K$ because $K$ is a cone.
Moreover, a slightly stronger property holds as well, for star-shaped $X$,
namely: if $\xi _1,\xi _2\in  X$ are such that $\xi _1\succeq\xi _2$ and if for some
linear map $L:\R^n\rightarrow \R^q$ it holds that $L\xi _1=L\xi _2$, then the sequences
$\{\xi ^i_1\},\{\xi ^i_2\}$ can be picked such that $L\xi ^i_1=L\xi ^i_2$ for all $i$;
this follows from the construction, since
$L(\xi ^i_1-\xi ^i_2)=\lambda ^iL(\xi _1-\xi _2)=0$.
For instance, $L$ might select those coordinates which belong in some subset
$I\subseteq \{1,\ldots ,n\}$.
This stronger property will be useful later, when we look at boundary points.
\er

The characterization will be in terms of a standard notion of tangent cone,
studied in nonsmooth analysis:
Let $S$ be a subset of a Euclidean space, and pick
any $\xi \in \So$.  The {\em tangent cone to $\So$ at} $\xi $ is the set
$\tanSx$
consisting of all
limits of the type
$\lim_{i\rightarrow \infty }\frac{1}{t_i}\left(\xi _i-\xi \right)$
such that $\xi _i \convS \xi $ and $t_i\searrow0$,
where ``$\xi _i \convS \xi $'' means that $\xi _i\rightarrow \xi $ as $i\rightarrow \infty $ and that
$\xi _i\in \So$ for all $i$.
Several properties of tangent cones are reviewed in the Appendix.
The main result in this section is as follows.

\bt{main-tangents}
The system~(\ref{cds}) is monotone if and only if, for all $\xi _1,\xi _2\in \Xo$:
\be{main-tangents-prop1}
\begin{array}{rcl}
\xi _1\succeq\xi _2,\,u_1\succeq u_2
\Rightarrow 
f(\xi _1,u_1)-f(\xi _2,u_2)\in \tanxdiffK
\end{array}
\ee
or, equivalently,
\be{main-tangents-prop2}
\xi _1-\xi _2\in \partial K ,\,u_1\succeq u_2
\Rightarrow 
 f(\xi _1,u_1)-f(\xi _2,u_2)\in \tanxdiffK
.
\ee
\ets

{\em Theorem~\ref{main-tangents} is valid} even if the relation
``$x_1 \succeq x_2$ iff $x_1-x_2 \in K$'' is defined {\em with respect to an
arbitrary closed set $K$}, not necesssarily a closed convex cone.
Our proof will not use the fact that $K$ is a a closed convex cone.
As a matter of fact, we may generalize even more.  Let us suppose that an
{\em arbitrary closed subset} $\Gamma \subseteq X\times  X$  has been given and we introduce the
relation, for $\xi _1,\xi _2\in X$:
\[
\xi _1\succeq\xi _2 \;\Leftrightarrow\; (\xi _1,\xi _2)\in \Gamma \,.
\]
We then define monotonicity just as in Definition~\ref{contrmondynsys}.
A particular case is $\Gamma =\Gamma (K)$, for a closed set $K$ (in particular, a convex
cone), with $(\xi _1,\xi _2)\in \Gamma (K)$ if and only if $x_1-x_2 \in K$.
Such an abstract setup is useful in the following situation: suppose that the
state dynamics are not necessarily monotone, but that we are interested in
output-monotonicity: if $u_1 \succeq u_2$ and $x_1 \succeq x_2$, then the
outputs satisfy $h(\phi (t,x_1,u_1)) \succeq h(\phi (t,x_2,u_2))$ for all $t$.
This last property is equivalent to the requirement that
$(\phi (t,x_1,u_1),\phi (t,x_2,u_2))\in \Gamma $, 
where $\Gamma $ is the set of all pairs of
states $(\xi _1,\xi _2)$ such that $h(\xi _1)\succeq h(\xi _2)$ in the output-value
order; note that $\Gamma $ is generally not of the form $\Gamma (K)$.
In order to provide a characterization for general $\Gamma $, we introduce
the system with state-space $X\times  X$ and input-value set
$\Inpt$ whose dynamics
\be{cdst}
\dot x = \ft(x,u)
\ee
are given, in block form using $x=(x_1,x_2)\in  X\times  X$ and $u=(u_1,u_2)\in \Inpt$,
as:
$\dot x_1=f(x_1,u_1)$,
$\dot x_2=f(x_2,u_2)$
(two copies of the same system, driven by the different $u_i$'s).
We will prove the following characterization, from which
Theorem~\ref{main-tangents} will follow as a corollary:

\bt{main-tangents-abstract}
The system~(\ref{cds}) is monotone if and only if, for all $\xi _1,\xi _2\in \Xo$:
\be{main-tangents-prop1-abstract}
\xi _1\succeq\xi _2 \;\;\mbox{and}\;\;u_1\succeq u_2
\;\Rightarrow \;
\ft(\xi ,u)\,\in \,\tanxG\,.
\ee
\ets

Returning to the case of orders induced by convex cones, we remark that
the conditions given in Theorem~\ref{main-tangents}
may be equivalently expressed in terms of a generalization,
to systems with inputs, of the property called {\em quasi-monotonicity}
(see for
instance~\cite{%
Kunze-Siegel,martin-sachs,Redheffer-Walter,Schneider-Vidyasagar,Volkmann,%
Walter}
and references therein): the system~(\ref{cds}) is monotone if and only if
\be{main-tangents-dual}
\begin{array}{rcl}
\xi _1\succeq\xi _2,\,u_1\succeq u_2,\,\zeta \in K^*,
\mbox{ and }
\ip{\zeta }{\xi _1}=\ip{\zeta }{\xi _2}
\\
\Rightarrow \ip{\zeta }{f(\xi _1,u_1)}\;\geq \;\ip{\zeta }{f(\xi _2,u_2)}
\end{array}
\ee
(it is enough to check this property for $\xi _1-\xi _2\in \partial K$),
where $K^*$ is the set of all $\zeta \in \R^n$ so that $\ip{\zeta }{k}\geq 0$ for all
$k\in  K$.
The equivalence follows from the elementary fact from convex analysis that,
for any closed convex cone $K$ and any element $p\in  K$, 
$\tanCpK$ coincides with the set of $v\in \R^n$ such that:
$
\ip{\zeta }{p}=0 \mbox{ and } \zeta \in K^* \,\Rightarrow \,\ip{v}{\zeta }\geq 0
$.
An alternative proof of Theorem~\ref{main-tangents} for the case of closed
convex cones $K$ should be possible by proving~(\ref{main-tangents-dual})
first, adapting the proofs and discussion in~\cite{Redheffer-Walter}.

Condition~(\ref{main-tangents-dual}) can be replaced by the conjunction of:
for all $\xi$ and all $u_1\succeq u_2$, $f(\xi,u_1)-f(\xi,u_2)\in K$, and for
all $u$, $\xi_1\succeq\xi_2$, and $\ip{\zeta}{\xi_1}=\ip{\zeta}{\xi_2}$,
$\ip{\zeta}{f(\xi_1,u)}\geq\ip{\zeta}{f(\xi_2,u)}$ (a similar separation is
possible in Theorem~\ref{main-tangents}).

The proofs of Theorems~\ref{main-tangents} and~\ref{main-tangents-abstract}
are given later.
First, we discuss 
the applicability of this test, and we develop several technical results.

We start by looking at a special case, namely $K = \R_{\geq 0}^n$ and
$K_u=\R_{\geq 0}^{m}$
(with $\banach_\Inputset=\R^m$).
Such systems are called {\em cooperative systems}.

The boundary points of $K$ are those points for which some coordinate is zero,
so ``$\xi _1-\xi _2\in \partial K$'' means that 
$\xi _1 \succeq \xi _2$
and
$\xi _1^i = \xi _2^i$ for at least one $i\in \{1,\ldots ,n\}$.
On the other hand, if $\xi _1 \succeq \xi _2$ and
$\xi _1^i = \xi _2^i$ for $i\in  I$ and 
$\xi _1^i > \xi _2^i$ for $i\in  \{1,\ldots ,n\}\setminus I$ ,
the tangent cone $\tanxdiffK$ consists of all those
vectors $v=(v_1,\ldots ,v_n)\in \R^n$
such that $v_i\geq 0$ for $i\in  I$ and $v_i$ is arbitrary in $\R$ otherwise.
Therefore, Property~(\ref{main-tangents-prop2}) 
translates into the following statement:
\be{cooperative2}
\begin{array}{rcl}
\xi _1 \succeq \xi _2 \textrm{ and } \xi _1^i = \xi _2^i \textrm{ and } u_1 \succeq
u_2 &\Rightarrow & \\
f^i (\xi _1,u_1) - f^i (\xi _2,u_2) &\geq& 0
\end{array}
\ee
holding
for all $i = 1,2, \ldots n$, all $u_1,u_2\in \Inputset$, and all
$\xi _1,\xi _2 \in \Xo$
(where $f^i$ denotes the $i$th component of $f$).
In particular, for systems with no inputs
$\dot {x} = f(x)$
one recovers the well-known characterization for cooperativity
(cf.~\cite{Smith}):
``$\xi _1 \succeq \xi _2$
and
$\xi _1^i = \xi _2^i$ implies
$f^i (\xi _1) \geq f^i (\xi _2)$''
must hold for all $i = 1,2, \ldots n$ and all $\xi _1,\xi _2 \in \Xo$.

When $X$ is strictly star-shaped, and in particular if $X$ is convex,
cf.\ Remark~\ref{convex-OK},
one could equally well require condition~(\ref{cooperative2}) to hold
for all $\xi _1,\xi _2 \in  X$.
Indeed, pick any $\xi _1 \succeq \xi _2$, and suppose that
$\xi _1^i = \xi _2^i$ for $i\in  I$ and 
$\xi _1^i > \xi _2^i$ for $i\in  \{1,\ldots ,n\}\setminus I$.
Pick sequences $\xi _1^k\rightarrow \xi _1$ and $\xi _2^k\rightarrow \xi _2$ so that, for all $k$,
$\xi _1^k,\xi _2^k\in \Xo$,
$\xi _1^k \succeq \xi _2^k$ and
$(\xi _1^k)^i = (\xi _2^k)^i$ for $i\in  I$
(this can be done by choosing an appropriate projection $L$ in
Remark~\ref{convex-OK}).
Since the property holds for elements in $\Xo$, we have that
$f^i (\xi _1^k,u_1)\geq f^i (\xi _2^k,u_2)$ for all $k=1,2,\ldots $ and all $i\in I$.
By continuity. taking limits as $k\rightarrow \infty $, we also have then that
$f^i (\xi _1,u_1)\geq f^i (\xi _2,u_2)$.
On the other hand, if $\Inputset$ also satisfies an approximability property,
then by continuity one proves similarly that it is enough to check the
condition~(\ref{cooperative2}) for $u_1,u_2$ belonging to the interior
$\Uo=\inter \Inputset$.
In summary, we can say that if $X$ and $\Inputset$ are both convex, then it is
equivalent to check condition~(\ref{cooperative2}) for elements in the sets
or in their respective interiors.

One can also rephrase the inequalities in terms of the partial derivatives of
the components of $f$.
Let us call a subset $S$ of an ordered Banach space {\em order-convex}
(``p-convex'' in~\cite{Smith}) if, for every $x$ and $y$ in $S$ with
$x\succeq y$ and every $0\leq \lambda \leq 1$, the element $\lambda  x+(1-\lambda )y$ is in $S$.
For instance, any convex set is order-convex, for all possible orders.
We have the following easy fact, which generalizes Remark 4.1.1
in~\cite{Smith}:

\bp{infinitesimal-derivs}
Suppose that $\Bu=\R^m$, $\Inputset$ satisfies an approximability property, and
both $\Xo$ and $\Uo=\inter \Inputset$ are order-convex
(for instance, these properties hold if both $\Xo$ and 
$\Inputset$
are convex).
Assume that $f$ is continuously differentiable.
Then, the system~(\ref{cds}) is cooperative if and only if the following
properties hold:
\beq
\label{infinitesimal-derivs-eq1}
\frac{\partial f^i}{\partial x^j} (x,u) \geq 0 \quad \forall \, x \in \Xo, \; 
\forall \, u\in \Uo,\; \forall \, i \neq j\\
\label{infinitesimal-derivs-eq2}
\frac{\partial f^i}{\partial u^j} (x,u) \geq 0 \quad \forall \, x \in X, \; \forall \, u
\in \Uo
\eeq
for all $i \in \{1,2, \ldots n \}$ and all $j \in \{1,2, \ldots m \}$.
\eps

\bpr
We will prove that these two conditions are equivalent to
condition~(\ref{cooperative2}) holding
for all $i = 1,2, \ldots n$, all $u_1,u_2\in \Uo$, and all
$\xi _1,\xi _2 \in \Xo$.
Necessity does not require the order-convexity assumption.
Pick any $\xi \in \Xo$, $u\in \Uo$, and pair $i\not= j$.
We take $\xi _1=\xi $, $u_1=u_2=u$, and $\xi _2(\lambda )=\xi +\lambda  e_j$, where
$e_j$ is the canonical basis vector having all coordinates $\not= j$ equal to zero
and its $j$th coordinate one, with $\lambda <0$ 
near enough to zero so that $\xi _2(\lambda )\in \Xo$.
Notice that, for all such $\lambda $,
$\xi _1\succeq\xi _2(\lambda )$ and $\xi _1^i = \xi _2(\lambda )^i$
(in fact, $\xi _1^\ell = \xi _2(\lambda )^\ell$ for all $\ell\not= j$).
Therefore condition~(\ref{cooperative2}) gives that
$f^i (\xi _1,u)\geq f^i (\xi _2(\lambda ),u)$
for all negative $\lambda \approx0$.
A similar argument shows that 
$f^i (\xi _1,u)\leq f^i (\xi _2(\lambda ),u)$
for all positive $\lambda \approx0$.
Thus $f^i (\xi _2(\lambda ),u)$ is increasing in a neighborhood of $\lambda =0$, 
and this implies Property~(\ref{infinitesimal-derivs-eq1}).
A similar argument establishes Property~(\ref{infinitesimal-derivs-eq2}).

For the converse,
as in~\cite{Smith}, we simply use the Fundamental Theorem of Calculus to write
$f^i(\xi _2,u_1)-f^i(\xi _1,u_1)$ as
$\int_0^1 \sum_{j=1}^n
\frac{\partial f^i}{\partial x^j} (\xi _1 + r(\xi _2-\xi _1),u_1 )(\xi _2^j-\xi _1^j)dr
$
and
$f^i(\xi _2,u_2)-f^i(\xi _2,u_1)$
as
$\int_0^1 \sum_{j=1}^m
\frac{\partial f^i}{\partial u^j} (\xi _2,u_2+r(u_2^j-u_1^j))(u_2^j-u_1^j)dr
$
for any $i = 1,2, \ldots n$, $u_1,u_2\in \Uo$, and $\xi _1,\xi _2 \in \Xo$.
Pick any $i\in \{1,2, \ldots n\}$, $u_1,u_2\in \Uo$, and $\xi _1,\xi _2 \in \Xo$, and
suppose that $\xi _1\succeq\xi _2$, $\xi _1^i=\xi _2^i$, and $u_1\succeq u _2$
We need to show that $f^i (\xi _2,u_2)\leq f^i (\xi _1,u_1)$.
Since the first integrand vanishes when $j=i$, and
also ${\partial f^i}/{\partial x^j}\geq 0$ and $\xi _2^j-\xi _1^j\leq 0$ for $j\not= i$, it follows that
$f^i(\xi _2,u_1)\leq f^i(\xi _1,u_1)$.
Similarly, the second integral formula gives us that
$f^i(\xi _2,u_2)\leq f^i(\xi _2,u_1)$, completing the proof.
\epr

For systems
without inputs,
Property~(\ref{infinitesimal-derivs-eq1}) is the well-known characterization
``$\frac{\partial f^i}{ \partial x^j} \geq 0 $ for all $i \neq j$'' of cooperativity.
Interestingly, the authors of~\cite{piccardi-rinaldi}
use this property, for systems as in~(\ref{cds}) but where 
inputs $u$ are seen as constant parameters,
as a definition of (parameterized)
cooperative systems, but monotonicity with
respect to time-varying inputs is not exploited there.
The terminology ``cooperative'' is motivated by this property: the different
variables $x^i$ have a positive influence on each other.

More general orthants can be treated by the trick used in Section 3.5
in~\cite{Smith}.  Any orthant $K$ in $\R^n$ has the form
$K^{(\varepsilon )}$, the set of all
$x\in \R^n$ so that $(-1)^{\varepsilon _i}x_i\geq 0$ for each $i=1,\ldots ,n$,
for some binary vector $\varepsilon =(\varepsilon _1,\ldots ,\varepsilon _n)\in \{0,1\}^n$.
Note that $K^{(\varepsilon )}=P\R_{\geq 0}^n$, where $P:\R^n\rightarrow \R^n$ is the linear mapping
given by the matrix $P={\rm diag}\,((-1)^{\varepsilon _1},\ldots ,(-1)^{\varepsilon _n})$.
Similarly, if the cone $K_u$ defining the order for $\Inputset$ is an orthant
$K^{(\delta )}$, we can view it as $Q\R_{\geq 0}^m$, for a similar map
$Q={\rm diag}\,((-1)^{\delta _1},\ldots ,(-1)^{\delta _m})$.
Monotonicity of $\dot x=f(x,u)$ under these orders is equivalent to
monotonicity of $\dot z=g(z,v)$, where 
$g(z,v)=Pf(Pz,Qv)$, under the already studied
orders given by $\R_{\geq 0}^n$ and $\R_{\geq 0}^{m}$.
This is because the change of variables $z(t)=Px(t)$, $v(t)=Qu(t)$ transforms
solutions of one system into the other (and viceversa), and both $P$ and $Q$
preserve the respective orders ($\xi _1\succeq\xi _2$ is equivalent to
$(P\xi _1)^i\geq (P\xi _2)^i$ for all $i\in \{1,\ldots ,n\}$,
and similarly for input values).
Thus we conclude:

\bc{infinitesimal-derivs-general-orthant}
Under the assumptions in Proposition~\ref{infinitesimal-derivs},
and for the orders induced from orthants $K^{(\varepsilon )}$ and $K^{(\delta )}$,
the system~(\ref{cds}) is monotone if and only if the following
properties hold for all $i \neq j$:
\be{infinitesimal-derivs-general-orthant-eq1}
(-1)^{\varepsilon _i+\varepsilon _j}\,
\frac{\partial f^i}{\partial x^j} (x,u) \geq 0 \quad \forall \, x \in \Xo, \; 
\forall \, u\in \Uo, %\; \forall \, i \neq j
\ee
\be{infinitesimal-derivs-general-orthant-eq2}
(-1)^{\varepsilon _i+\delta _j}\,
\frac{\partial f^i}{\partial u^j} (x,u) \geq 0 \quad \forall \, x \in X, \; \forall \, u
\in \Uo
\ee
for all $i \in \{1,2, \ldots n \}$ and all $j \in \{1,2, \ldots m \}$.
\ec

Graphical characterizations of monotonicity with respect to orthants are
possible; see~\cite{part2} for a discussion.  The conditions amount to asking
that there should not be any negative (non-oriented) loops in the incidence
graph of the system.

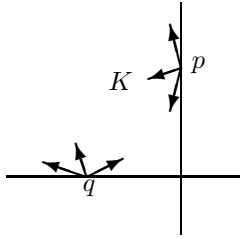
\begin{figure}[htl]
\setlength{\unitlength}{1200sp}%
\begin{center}
\begin{picture}(4824,4824)(3289,-7573)
\thinlines
\put(6901,-7561){\line( 0, 1){4800}}
\put(8101,-6361){\line(-1, 0){4800}}
\thicklines
\put(4951,-6361){\vector( 2, 1){750}}
\put(4951,-6361){\vector(-1, 3){225}}
\put(4951,-6361){\vector(-3, 1){900}}
\put(6901,-4111){\vector(-1,-4){225}}
\put(6901,-4111){\vector(-3,-1){675}}
\put(6901,-4111){\vector(-1, 4){225}}
\put(5401,-4561){$K$}
\put(7126,-4111){$p$}
\put(4876,-6661){$q$}
\end{picture}
\end{center}
\caption{Example of cone and tangents}
\label{fig-boundary-example}
\end{figure}

Let us clarify the above definitions and notations with an example.
We consider the partial order $\succeq$ obtained by letting
$K=\R_{\leq 0} \times \R_{\geq 0}$.
Using the previous notations, we can write this as
$K=K^{(\varepsilon )}$, where $\varepsilon =(1,0)$.
We will consider the input space $\Inputset=\R_{\geq 0}$, with the 
standard ordering in $\R$ (i.e., $K_u=\R_{\geq 0}$, or $K_u=K^{(\delta )}$ with
$\delta =(0)$).
Observe that the boundary points of the cone $K$ are those points of the forms
$p=(0,a)$ or $q=(-a,0)$, for some $a \geq 0$,
and the tangent cones are respectively 
$\tanCpK=\R_{\leq 0}\times \R$
and $\tanCpK=\R\times \R_{\geq 0}$,
see Fig.~\ref{fig-boundary-example}.
Under the assumptions of Corollary~\ref{infinitesimal-derivs-general-orthant},
a system is monotone with respect to these orders if and only if the following
four inequalities hold everywhere:
\[
\frac{\partial f^1}{\partial x^2} \leq  0\,,\;
\frac{\partial f^2}{\partial x^1} \leq  0\,,\;
\frac{\partial f^1}{\partial u} \leq 0  \,,\;
\frac{\partial f^2}{\partial u} \geq 0  \,.
\]
A special class of systems of this type is afforded by systems as follows:
\begin{equation}
\label{nlsys}
\left .
\begin{array}{l}
\dot {x}^1 = - u \theta _1(x^1) + \theta _2 ( 1- x^1 - x^2 )  \\
\dot {x}^2 = u \theta _3 ( 1 -x^1 - x^2 ) - \theta _4 ( x^2 ) 
\end{array}
\right \} = f(x^1,x^2,u) 
\end{equation}
where the functions $\theta _i$ have strictly positive derivatives and satisfy
$\theta _i(0)=0$. The system is regarded as evolving on the triangle
$
X=\Delta   :=\{ [x^1,x^2]: x^1 \geq 0, \; x^2 \geq 0, x^1+x^2 \leq 1 \}
$,
which is easily seen to be invariant for the dynamics.
Such systems arise after restricting to the affine subspace
$x^1+x^*+x^2=1$ and eliminating the variable $x^*$ in the set of three
equations
$\dot {x}^1 = - u \theta _1(x^1) + \theta _2 (x^*),
\dot x^* = u \theta _1(x^1) - \theta _2 (x^*) - u \theta _3 (x^*) + \theta _4 ( x^2 ),
\dot {x}^2 = u \theta _3 (x^*) - \theta _4 ( x^2 )$,
and they model an important component of cellular processes,
see e.g.~\cite{ferrell,kholodenko} and the discussion in 
Section~\ref{An Application}.
(The entire system, before eliminating $x^*$, can also be shown directly to be
monotone, by means of the change of coordinates $y_1=x^1$, $y_2=x^1+x^*$,
$y_3=x^1+x^*+x^2$.  As such, it, and analogous higher-dimensional signaling
systems, are ``cooperative tridiagonal systems'' for which a rich theory
of stability exists; this approach will be discussed in future work.) 
The following fact is immediate from the above discussion:
\bl{mapk-basic-piece-is-monotone}
The system~(\ref{nlsys}) is monotone with respect to the given orders.
\el

\br{competitive}
One may also define {\em competitive systems}
as those for which
$u_1 \succeq u_2$ and $x_1 \succeq x_2$ imply
$\phi (t,x_1,u_1) \succeq \phi (t,x_2,u_2)$
for $t \leq 0$.
Reversing time, one obtains the characterization:
``$\xi _1-\xi _2\in \partial K$ and $u_1\succeq u_2$
$\Rightarrow $ $f(\xi _2,u_2)-f(\xi _1,u_1)\in \tanxdiffK$''
or, for the special case of the positive orthant,
$ \frac{\partial f^i}{\partial x^j} (x,u) \leq 0$ for all $x \in X$ and all
$u\in \Inputset$ ($i \neq j$)
together with
$\frac{\partial f^i}{\partial u^j} (x,u) \leq 0$ for all $x \in X$ and all
$u\in \Inputset$
for all $i \in \{1,2, \ldots n \}$ and all $j \in \{1,2, \ldots m \}$.
\er

We now return to the proof of Theorem~\ref{main-tangents}.

\bl{int-fi}
The set $\Xo$ is forward invariant for~(\ref{cds}), i.e., for each
$\xi \in \Xo$ and each $u\in \Inpf$, $\phi (t,\xi ,u)\in \Xo$ for all $t\geq 0$.
\els

\bpr
Pick any $\xi \in \Xo$, $u\in \Inpf$, and $t_0\geq 0$.
Viewing~(\ref{cds}) as a system defined on an open set of states $\widetilde X$
which contains $X$, we consider the mapping $\alpha :\Xo\rightarrow \banach$ given by
$\alpha (x)=\phi (t_0,x,u)$ (with the same $u$ and $t_0$).
The image of $\alpha $ must contain a neighborhood $W$ of $\xi '=\alpha (\xi )$;
see e.g.\ Lemma~4.3.8 in~\cite{mct}.
Thus, $W\subseteq  X$, which means that $\xi '\in \inter X$, as desired.
\epr

\br{rem-converse-fi}
The converse of Lemma~\ref{int-fi} is also true, namely, if $\dot x=f(x,u)$ is a
system defined on some neighborhood $\widetilde X$ of $X$ and if $\Xo=\inter X$ is
forward invariant under solutions of this system, 
then $X$ is itself invariant.
To see this, pick any $\xi \in  X$ and a sequence $\xi ^i\rightarrow \xi $ of elements of $\Xo$.
For any $t$, $i$, and $u$, $\phi (t,\xi ^i,u)\in \Xo$, so
$\phi (t,\xi ,u)=\lim_{i\rightarrow \infty }\phi (t,\xi ^i,u)\in \clos\Xo=X$.
\er

We introduce 
the closed set
$\Inpt$
consisting of all $(u_1,u_2)\in {\cal U}\times  {\cal U}$
such that $u_1\succeq u_2$.
We denote by $\Inpft$ the set of all possible inputs 
to the composite system~(\ref{cdst})
i.e., the set of all Lebesgue-measurable locally essentially bounded functions
$u:[0,\infty )\rightarrow \Inpt$.
Since by Lemma~\ref{int-fi} the interior $\Xo$ of $X$ is forward invariant
for~(\ref{cds}), it holds that $\fit(t,\xi ,u)$ belongs to $\Xo\times  \Xo$
whenever $\xi \in  \Xo\times  \Xo$ and $u\in \Inpft$.

Observe that the definition of monotonicity amounts to the requirement that:
{\em for each $\xi \in \Gamma $, and each $u\in \Inpft$, the solution $\fit(t,\xi ,u)$
of~(\ref{cdst}) with initial condition $x(0)=\xi $ belongs to $\Gamma $ for all
$t\geq 0$} (forward invariance of $\Gamma $ with respect to~(\ref{cdst})).
Also, the set
$
\Gamma _0:= \Gamma  \bigcap  \left(\Xo\times \Xo\right)
$
is closed relative to $\Xo \times \Xo$. 
The following elementary remark will be very useful:
\bl{monotone=gamma0fi}
The system~(\ref{cds}) is monotone if and only if
the set $\Gamma _0$ is forward invariant for the system~(\ref{cdst}) restricted to
$\Xo\times \Xo$.
\els

\bpr
We must show that monotonicity is the same as:
``$\xi \in \Gamma _0$ and $u\in \Inpft$ $\Rightarrow $ $\fit(t,\xi ,u)\in \Gamma _0$
for all $t \geq 0$.''
Necessity is clear, since if the system is monotone then 
$\fit(t,\xi ,u)\in \Gamma $ holds for all $\xi \in \Gamma \supseteq \Gamma _0$ and all $t\geq 0$,
and we already remarked that $\fit(t,\xi ,u)\in \Xo\times \Xo$ whenever $\xi \in \Gamma _0$.
Conversely, 
pick any $\xi \in \Gamma $.
The approximability hypothesis provides a sequence $\{\xi ^i\}\subseteq \Gamma _0$
such that $\xi ^i\rightarrow \xi $ as $i\rightarrow \infty $.
Fix any $u\in \Inpft$ and any $t\geq 0$.
Then $\fit(t,\xi ^i,u)\in \Gamma _0\subseteq \Gamma $ for all $i$, so taking limits and using
continuity of $\fit$ on initial conditions gives that $\fit(t,\xi ,u)\in \Gamma $, as
required.
\epr

\bl{two-tangent-chars}
For any $\xi =(\xi _1,\xi _2)\in \Gamma _0$ and any $u=(u_1,u_2)\in \Inpt$, the following
three properties are equivalent:
\be{f-f}
f(\xi _1,u_1)-f(\xi _2,u_2)\,\in \, \tanxdiffK
\ee
\be{f2}
\ft(\xi ,u)\,\in \,\tanxGo
\ee
\be{f3}
\ft(\xi ,u)\,\in \,\tanxG\,.
\ee
\els

\bpr
Suppose that~(\ref{f-f}) holds, so there are sequences $t_i\searrow0$ and
$\{\eta ^i\}\subseteq  K$ such that $\eta ^i\rightarrow \xi _1-\xi _2$ and
\be{ttc-eq1}
\frac{1}{t_i}\left(\eta ^i-(\xi _1-\xi _2)\right) \;\rightarrow \;
f(\xi _1,u_1)-f(\xi _2,u_2)
\ee
as $i \rightarrow \infty$.
Since $\Xo$ is open, the solution $x(t)=\phi (t,\xi _1,\ubar)$ 
of $\dot x=f(x,\ubar)$ with input $\ubar\equiv u_1$ and initial condition $x(0)=\xi _1$
takes values in $\Xo$ for all sufficiently small $t$.
Thus, restricting to a subsequence, we may without loss of generality assume
that $\xi ^i_1:=x(t_i)$ is in $\Xo$ for all $i$.
Note that, by definition of solution,
(a) $\fract{1}{t_i}(\xi ^i_1-\xi _1)  \rightarrow 
f(\xi _1,u_1)$
as $i\rightarrow \infty $, and subtracting~(\ref{ttc-eq1}) from this we obtain that
(b) $\fract{1}{t_i} (\xi ^i_2-\xi _2) \rightarrow f(\xi _2,u_2)$
as $i\rightarrow \infty $,
with $\xi ^i_2:=\xi ^i_1-\eta ^i$.
Since $\xi ^i_1\rightarrow \xi _1$ and $\eta ^i\rightarrow \xi _1-\xi _2$ as $i\rightarrow \infty $, 
the sequence $\xi ^i_2$ converges to $\xi _2\in \Xo$.
Using once again that $\Xo$ is open, we may assume without loss of generality
that $\xi ^i_2\in \Xo$ for all $i$.
Moreover, $\xi ^i_1-\xi ^i_2=\eta ^i\in  K$, i.e., 
$\xi ^i:=(\xi ^i_1,\xi ^i_2)\in \Gamma $ 
for all $i$,
which means that $\xi ^i$ is in $\Gamma _0$ for all $i$, and, from
the previous considerations,
(c) $\fract{1}{t_i}(\xi ^i-\xi )  \rightarrow 
(f(\xi _1,u_1),f(\xi _2,u_2))$
as $i\rightarrow \infty $,
so that Property~(\ref{f2}) is verified.
Since $\Gamma _0\subseteq \Gamma $, also Property~(\ref{f3}) holds.

Conversely, suppose that Property~(\ref{f3}) holds.
Then there are
sequences $t_i\searrow0$ and $\xi ^i:=(\xi ^i_1,\xi ^i_2)\in \Gamma $
with $\xi ^i\rightarrow \xi $ such
that~(c) holds.
Since $\xi \in {\cal V}\times {\cal V}$, we may assume without loss of generality that
$\xi ^i\in \Gamma _0$ for all $i$, so that we also have Property~(\ref{f2}).
Coordinatewise, we have
both~(a) and~(b), which subtracted
and defining $\eta ^i:=\xi ^i_1-\xi ^i_2$ give~(\ref{ttc-eq1}); this establishes
Property~(\ref{f-f}).
\epr

\noindent{\bf Proofs of Theorems~\protect{\ref{main-tangents}}
and~\protect{\ref{main-tangents-abstract}}}

Suppose that the system~(\ref{cds}) is monotone, and fix any input-value
pair $u^0=(u_1^0,u_2^0)\in \Inpt$.
Lemma~\ref{monotone=gamma0fi} says that the set $\Gamma _0$ is forward invariant
for the system~(\ref{cdst}) restricted to $\Xo\times \Xo$.
This implies, in particular, that every solution of the differential equation
$
\dot x=\ft(x,u^0)
$
with $x(0)\in \Gamma _0$ remains in $\Gamma _0$ for all $t\geq 0$ (where we think of
$u^0$ as a constant input).
We may view this differential equation as a (single-valued) differential
inclusion $\dot x\in  F(x)$ on $\Xo\times \Xo$, where $F(\xi )=\{\ft(\xi ,u^0)\}$, for which
the set $\Gamma _0$ is strongly invariant.
Thus, Theorem~\ref{main-theorem-invariance} in the Appendix implies that
$F(\xi )\subseteq \tanxGo$ for all $\xi \in \Gamma _0$.
In other words, Property~(\ref{f2}), or equivalently
Property~(\ref{f3}) holds, at all $\xi \in \Gamma _0$, for the given
$u=u^0$.  Since $u^0$ was an arbitrary element of $\Inpt$,
Property~(\ref{main-tangents-prop1-abstract}) follows.
By Lemma~\ref{two-tangent-chars},
$f(\xi _1,u^0_1)-f(\xi _2,u^0_2)\in \tanxdiffK$ for all $(\xi _1,\xi _2)\in \Gamma _0$ and this
$u^0$.  So Property~(\ref{main-tangents-prop1}) also follows.

Conversely, suppose that~(\ref{main-tangents-prop1}) holds
or~(\ref{main-tangents-prop1-abstract}) holds.
By Lemma~\ref{two-tangent-chars}, we know that 
Property~(\ref{f2})
holds for all $(\xi _1,\xi _2)\in \Gamma _0$ and all $(u_1,u_2)\in \Inpt$.
To show monotonicity of the system~(\ref{cds}), we need to prove that $\Gamma _0$ is
invariant for the system~(\ref{cdst}) when restricted to $\Xo\times \Xo$.
So pick any $\xi ^0\in \Gamma _0$, any $u^0\in \Inpft$, and any $t^0>0$; we must prove
that $\fit(t^0,\xi ^0,u^0)\in \Gamma _0$.
The input function $u^0$ being locally bounded means that there is some
compact subset $C\subseteq \Inputset$ such that $u(t)$ belongs to the compact subset
$\Inpt_C=\Inpt\bigcap  C\times  C$ of $\banach_\Inputset\times  \banach_\Inputset$, for (almost) all
$t\in [0,t^0]$.
We introduce the following compact-valued, locally bounded, and locally
Lipschitz set-valued function:
$
F_C(\xi ):=
\{\ft(\xi ,u)\st u\in \Inpt_C \}
$
on $\Xo\times \Xo$.
We already remarked that Property~(\ref{f-f}) holds, i.e.,
$\{\ft(\xi ,u)\st u\in \Inpt\}\subseteq \tanxGo$,
for all $(\xi _1,\xi _2)\in \Gamma _0$, so it is true in particular that
$F_C(\xi )\subseteq \tanxGo$.
Thus, Theorem~\ref{main-theorem-invariance} in the Appendix implies that
$\Gamma _0$ is strongly invariant with respect to $F_C$.
Thus, since $x(\cdot )=\fit(\cdot ,\xi ^0,u^0)$ restricted to $[0,t^0]$
satisfies $\dot x\in  F_C(x)$, we conclude that $x(t^0)\in \Gamma _0$, as required.

Finally, we show that~(\ref{main-tangents-prop1})
and~(\ref{main-tangents-prop2}) are equivalent.
Since~(\ref{main-tangents-prop2}) is a particular case
of~(\ref{main-tangents-prop1}), we only need to verify that
$f(\xi _1,u_1)-f(\xi _2,u_2)\in \tanxdiffK$ when $\xi _1-\xi _2\in \inter K$.
This is a consequence of the general fact that $\tanSx=\R^n$ whenever
$\xi $ is in the interior of a set $S$.
\qed

\section{Cascades of monotone systems}
\label{Cascades of monotone systems}

Cascade structures with triangular form 
\be{triangular}
\begin{array}{rcl}
\dot {x}_1 & = & f_1 (x_1 , x_2, \ldots , x_N,u ) \\
\dot {x}_2 & = & f_2 (x_2 , \ldots , x_N ,u) \\
\vdots &  & \vdots \\
\dot {x}_N & = & f_N (x_N,u)  
\end{array}
\ee
are of special interest. A simple sufficient condition for 
monotonicity of systems (\ref{triangular}) is as follows.

\bp{cascaderesult}
Assume that there exist positivity cones $K_1, K_2, \ldots , K_{N+1}$ (of suitable
dimensions) so that each of the
$x_i$-subsystems in (\ref{triangular}) is a controlled
monotone dynamical system with respect to the $K_i$-induced partial
order (as far as states are concerned) and with respect to the 
$K_{i+1},\ldots , K_{N+1}$-induced 
partial orders as far as inputs are concerned.
Then, the overall cascaded interconnection (\ref{triangular}) is monotone with
respect to the order induced by the positivity cone 
$K_1 \times K_2 \times \ldots  \times K_N$ on states and $K_{N+1}$ on inputs.
\eps

\bpr We first prove the result for the case $N=2$:
$\dot {x}_1 = f_1 (x_1,x_2,u)$,
$\dot {x}_2 = f_2 (x_2,u)$.
Let $\succeq_1$ and $\succeq_2$ 
be the partial orders induced by the cones $K_1$,$K_2$ and $\succeq_u$ on
inputs.
Pick any two inputs $u^a \succeq_u u^b$.
By hypothesis we have, for each two states
$\xi ^a=(\xi ^a_1,\xi ^a_2)$ and $\xi ^b=(\xi ^b_1,\xi ^b_2)$, that
$\xi ^a_2 \succeq_2 \xi ^b_2$ implies
$\phi _2 (t, \xi ^a_2,u^a) \succeq_2 \phi _2(t,\xi ^b_2,u^b)$
for all $t \geq 0$
as well as, for all functions $x^a_2$ and $x^b_2$ that
$\xi ^a_1 \succeq_1 \xi ^b_1$
and $x^a_2  \succeq_2 x^b_2$ implies
$\phi _1 (t,\xi ^a_1,x_2^a,u^a)\succeq_1 \phi _1 (t,\xi ^b_1,x_2^b,u^b)$
for all $t \geq 0$.
Combining these, and defining $K:=K_1 \times K_2$ and letting $\succeq$ denote the
corresponding partial order, 
we conclude that
$\xi ^a \succeq \xi ^b$
implies
$\phi  ( t,  \xi ^a, u^a  ) \succeq \phi  (t,\xi ^b,u^b)$
for all $t \geq 0$.
The proof for arbitrary $N$ follows by induction.
\epr

\section{Static Input/State and Input/Output Characteristics}
\label{Static Input/State and Input/Output Characteristics}

A notion of ``Cauchy gain'' was introduced in~\cite{cauchy} to quantify
amplification of signals in a manner useful for biological applications.
For monotone dynamical systems satisfying an additional property, it
is possible to obtain tight estimates of Cauchy gains.  
This is achieved by showing that the output values $y(t)$ corresponding to an
input $u(\cdot )$ are always ``sandwiched'' in between the outputs corresponding to
two constant inputs which bound the range of $u(\cdot )$.
This additional property motivated our looking at monotone
systems to start with; we now start discussion of that topic.

\bd{ISchar}
We say that a controlled dynamical system~(\ref{cds})
is endowed with the \emph{static Input/State characteristic}
\[
k_x (\cdot ):\Inputset \rightarrow  X
\]
if for each constant input $u(t) \equiv \ubar$ there exists a (necessarily unique)
globally
asymptotically stable equilibrium $k_x (\ubar)$. 
For systems with an output map $y = h(x)$, we also define the 
\emph{static Input/Output characteristic} as
$k_y (\ubar) := h(k_x(\ubar))$,
provided that an Input/State characteristic exists
and that $h$ is continuous.
\ed
The paper~\cite{piccardi-rinaldi} (see also~\cite{muratori-rinaldi}
for linear systems) provides very useful results which can be used to
show the existence of I/S characteristics, for cooperative systems
with scalar inputs and whose state space is the positive orthant, and
in particular to the study of the question of when $k_x (\ubar)$ is
strictly positive.

\br{monotone-char}
Observe that, if the system~(\ref{cds}) is monotone and it admits a 
static Input/State characteristic $k_x$, then $k_x$ must be nondecreasing with
respect to the orders in question:
$\ubar\succeq \vbar$ in $\Inputset$ implies $k_x(\ubar)\succeq k_x(\vbar)$.
Indeed, given any initial state $\xi $, monotonicity says
that $\phi (t,\xi ,u)\succeq\phi (t,\xi ,v)$ for all $t$, where $u(t)\equiv \ubar$ and
$v(t)\equiv \vbar$.  Taking limits as $t\rightarrow \infty $ gives the desired conclusion.
\er

\br{c0-char}
(Continuity of $k_x$) \ 
Suppose that for a system~(\ref{cds}) there is a map
$k_x:\Inputset \rightarrow  X$
with the property that $k_x(\ubar)$ is the unique steady state of the
system $\dot x=f(x,\ubar)$ (constant input $u\equiv \ubar$).
When $k_x(\ubar)$ is a globally asymptotically stable state for 
$\dot x=f(x,\ubar)$, as is the case for I/S characteristics, it follows that
the function $k_x$ must be continuous, see Proposition~\ref{cics} below.
However, continuity is always true provided only that $k_x$ be locally bounded,
i.e.\ that $k_x(V)$ is a bounded set whenever $V\subseteq \Inputset$ is compact.
This is because $k_x$ has a closed graph, since
$k_x(\ubar)=\xbar$ means that $f(\xbar,\ubar)=0$, and any locally bounded map
with a closed graph (in finite-dimensional spaces) must be continuous.
(Proof: suppose that $\ubar_i\rightarrow \ubar$, and consider the sequence
$\xbar_i=k_x(\ubar_i)$; by local boundedness, it is only necessary to prove
that every limit point of this sequence equals $k_x(\ubar)$.  
So suppose that
$\xbar_{i_j}\rightarrow \xbar'$; then $(\ubar_{i_j},\xbar_{i_j})\rightarrow (\ubar,\xbar')$, so by
the closedness of the graph of $k_x$ we know that $(\ubar,\xbar')$ belongs to
its graph, and thus $\xbar'=\xbar$, as desired.)  Therefore, local
boundedness, and hence continuity of $k_x$, would follow if one knows that
$k_x$ is monotone, so that $k([a,b])$ is always bounded, even if the stability
condition does not hold, at least if the order is ``reasonable'' enough, as in
the next definition.
Note that $k_y$ is continuous whenever $k_x$ is, since the output map $h$ has
been assumed to be continuous.
\er

Under weak assumptions, existence of a static Input/State characteristic
implies that the system behaves well with respect to arbitrary bounded inputs
as well as inputs that converge to some limit.
For convenience in stating results along those lines, we introduce the
following terminology:
The order on $X$ is {\em bounded} if the following two properties hold:
(1)
For each bounded subset $S\subseteq  X$, there exist two elements $a,b\in \banach$ such
that $S\subseteq [a,b]=\{ x \in X: a \preceq x \preceq b\}$,
and
(2)
For each $a,b\in \banach$, the set $[a,b]$ is bounded.
Boundedness is a very mild assumption.
In general, Property 1 holds if (and only if) $K$ has a nonempty interior,
and Property 2 is a consequence of $K\bigcap -K=\{0\}$.
(The proof is an easy exercise in convex analysis.)

\bp{bibs}
Consider a monotone system~(\ref{cds}) which is endowed with a static
Input/State characteristic,
and suppose that the order on the state space $X$ is bounded.
Pick any input $u$ all whose values $u(t)$ lie in some interval
$[c,d]\subseteq \Inputset$. 
(For example, $u$ could be any bounded input, if $K$ is an orthant in $\R^n$,
or more generally if the order in $\Inputset$ is bounded.)
Let $x(t)=\phi (t,\xi ,u)$ be any
trajectory of the system corresponding to this control.
Then $\{x(t), t\geq 0\}$ is a bounded subset of $X$.
\eps

\bpr
Let $x_1(t)=\phi (t,\xi ,d)$, so $x_1(t)\rightarrow k_x(d)$ as $t\rightarrow \infty $ and, in
particular, $x_1(\cdot )$ is bounded; so (bounded order), there is some
$b\in \banach$ such that $x_1(t)\preceq b$ for all $t\geq 0$.
By monotonicity,
$
x(t)=\phi (t,\xi ,u)\preceq \phi (t,\xi ,d) = x_1(t) \preceq b
$
for all $t\geq 0$. 
A similar argument using the lower bound $c$ on $u$ shows that there is some
$a\in \banach$ such that $a \preceq x(t)$ for all $t$.
Thus $x(t)\in [a,b]$ for all $t$, which implies, again appealing to the bounded
order hypothesis, that $x(\cdot )$ is bounded.
\epr

Certain standard facts concerning the robustness of stability will be useful.
We collect the necessary results in the next statements, for easy reference.

\bp{cics}
If~(\ref{cds}) is a monotone system which is endowed with a static
Input/State characteristic $k_x$, then  $k_x$ is a continuous map.
Moreover for each $\ubar\in \Inputset$, $\xbar=k_x(\ubar)$,
the following properties hold:
\ben
\item
For each neighborhood $P$ of $\xbar$ in $X$
there exist a neighborhood $P_0$ of $\xbar$ in $X$, and a
neighborhood $Q_0$ of $\ubar$ in $\Inputset$, such that
$\phi (t,\xi ,u)\in  P$ for all $t\geq 0$, all $\xi \in  P_0$, and all
inputs $u$ such that $u(t)\in  Q_0$ for all $t\geq 0$.
\item
If in addition the order on the state space $X$ is bounded,
then, for each input $u$ all whose values $u(t)$ lie in some interval
$[c,d]\subseteq \Inputset$ and with the property that $u(t)\rightarrow \bar u$, and all initial
states $\xi \in  X$, necessarily $x(t)=\phi (t,\xi ,u)\rightarrow \xbar$ as $t\rightarrow \infty $.
\een
\eps

\bpr
Consider any trajectory $x(t)=\phi (t,\xi ,u)$ as in Property 2.
By Proposition~\ref{bibs}, we know that there is some compact $C\subseteq \banach$
such that $x(t)\in  C$ for all $t\geq 0$.
Since $X$ is closed, we may assume that $C\subseteq  X$.
We are therefore in the following situation:
the autonomous system $\dot x=f(x,\ubar)$ admits $\xbar$ as a globally
asymptotically stable equilibrium (with respect to the state space $X$)
and the trajectory $x(\cdot )$ remains in a compact subset of the domain of
attraction (of $\dot x=f(x,\ubar)$ seen as a system on an open subset of
$\banach$ which contains $X$).
The ``converging input converging state'' property then holds for this
trajectory
(see~\cite{cics}, Theorem 1, for details).
Property 1 is a consequence of the same results.
(As observed to the authors by German Enciso, the CICS property can be also
verified as a consequence of ``normality'' of the order in the state space.)
The continuity of $k_x$ is a consequence of Property 1.
As discussed in Remark~\ref{c0-char}, we only need to show that $k_x$ is
locally bounded, for which it is enough to show that for each $\ubar$ there is
some neighborhood $Q_0$ of $\ubar$ and some compact subset $P$ of $X$ such that
$k_x(\mu )\in  P$ for all $\mu \in  Q_0$.
Pick any $\ubar$, and any compact neighborhood $P$ of $\xbar=k_x(\ubar)$.
By Property 1, there exist a neighborhood $P_0$ of $\xbar$ in $X$, and a
neighborhood $Q_0$ of $\ubar$ in $\Inputset$, such that
$\phi (t,\xi ,u_\mu )\in  P$ for all $t\geq 0$ whenever $\xi \in  P_0$ and $u_\mu (t)\equiv \mu $
with $\mu \in  Q_0$.
In particular, this implies that $k_x(\mu )=\lim_{t\rightarrow \infty }\phi (t,\xbar,u_\mu )\in  P$,
as required.
\epr

\bc{cascades}
Suppose that the system $\dot x=f(x,u)$ with output $y=h(x)$ is monotone and has 
static Input/State and Input/Output characteristics $k_x$, $k_y$, and that 
the system $\dot z=g(z,y)$ (with input value space equal to the output value
space of the first system and 
 the orders induced by the same positivity cone holding in the two spaces)
has a static Input/State characteristic $k_z$,
it is monotone, and the order on its state space $Z$ is bounded.
Assume that the order on outputs $y$ is bounded,
Then the cascade system
\beqn
\dot x&=&f(x,u)\,,\quad y=h(x)\\
\dot z&=&g(z,y)
\eeqn
is a monotone system which admits the static Input/State characteristic 
$\widetilde k(\ubar)=(k_x(\ubar),k_z(k_y(\ubar)))$.
\ecs
\bpr
Pick any $\ubar$.
We must show that $\widetilde k(\ubar)$ is a globally asymptotically
stable equilibrium (attractive and Lyapunov-stable)
of the cascade.
Pick any initial state $(\xi ,\zeta )$ of the composite system,
and let $x(t)=\phi _x(t,\xi ,\ubar)$ (input constantly equal to $\ubar$),
$y(t)=h(x(t))$, and $z(t)=\phi _z(t,\zeta ,y)$.
Notice that $x(t)\rightarrow \xbar$ and
$y(t)=h(x(t))\rightarrow \ybar=k_y(\ubar)$, so viewing $y$ as an input to
the second system and using Property 2 in Proposition~\ref{cics},
we have that $z(t)\rightarrow \zbar=k_z(k_y(\ubar))$.  This establishes attractivity.
To show stability, pick any neighborhoods $P_x$ and $P_z$ of $\xbar$ and
$\zbar$ respectively.
By Property 1 in Proposition~\ref{cics}, there are neighborhoods $P_0$ and
$Q_0$ such that $\zeta \in  P_0$ and $y(t)\in  Q_0$ for all $t\geq 0$ imply
$\phi _z(t,\zeta ,y)\in P_z$ for all $t\geq 0$.
Consider $P_1:=P_x\bigcap  h^{-1}(Q_0)$, 
which is a neighborhood of $\xbar$, and
pick any neighborhood $P_2$ of $\xbar$ with the property that
$\phi (t,\xi ,\ubar)\in  P_1$ for all $\xi \in  P_2$ and all $t\geq 0$ (stability of the
equilibrium $\xbar$).
Then, for all $(\xi ,\zeta )\in P_2\times  P_0$, $x(t)=\phi _x(t,\xi ,\ubar)\in  P_1$ (in
particular, $x(t)\in  P_x$) for all
$t\geq 0$, so $y(t)=h(x(t))\in  Q_0$, and hence also
$z(t)=\phi _z(t,\zeta ,y)\in  Q_z$ for all $t\geq 0$.
\epr

In analogy to what usually done for autonomous dynamical systems,
we define the $\Omega $-limit set of any function $\alpha :[0,\infty )\rightarrow A$, where $A$ is a
topological space (we will apply this to state-space solutions and to outputs)
as
$\Omega [\alpha ]:= \{a\in  A \st \exists \; t_k \rightarrow + \infty  \textrm{ s.t. }
\lim_{k \rightarrow + \infty} \alpha (t_k) = a  \}$
(in general, this set may be empty).
For inputs $u\in \Inpf$, we also introduce the sets
${\cal L}_{\leq }[u]$ (respectively, ${\cal L}_{\geq }[u]$)
consisting of all $\mu \in \Inputset$
such that there are $t_k \rightarrow + \infty $  and
$\mu _k \rightarrow \mu $ ($k \rightarrow + \infty $) with $\mu _k \in \Inputset$ so that
$u(t)\succeq \mu _k$ 
(respectively $\mu _k \succeq u(t)$)
for all $t\geq t_k$.
These notations are motivated by the following special case:
Suppose that we consider a {\em SISO\/} (single-input single-output) system, 
by which we mean a system for which $\Bu=\R$ and $\By=\R$, taken with the
usual orders.
Given any scalar bounded input $u(\cdot )$, we denote
$u_{\inf}:=\liminf_{t \rightarrow + \infty} u(t)$
and
$u_{\sup} \,:= \;\limsup_{t \rightarrow + \infty} u(t)$.
Then, $u_{\inf}\in {\cal L}_{\leq }[u]$
and $u_{\sup}\in {\cal L}_{\geq }[u]$, as follows by definition of lim inf and lim sup.
Similarly, both
$\liminf_{t \rightarrow + \infty} y(t)$ and
$\limsup_{t \rightarrow + \infty} y(t)$ belong to $\Omega [y]$, for any output $y$.

\bp{asybehaviour-general}
Consider a monotone system (\ref{cds}), with static I/S and I/O
characteristics $k_x$  and $k_y$.
Then, for each initial condition $\xi $ and each input $u$, the solution
$x(t)=\phi (t,\xi ,u)$ and the corresponding output $y(t)=h(x(t))$ satisfy:
\beqn
k_x\left({\cal L}_{\leq }[u]\right) \;\preceq\; \Omega [x] &\preceq&
                                             k_x\left({\cal L}_{\geq }[u]\right)\\
k_y\left({\cal L}_{\leq }[u]\right) \;\preceq\; \Omega [y] &\preceq&
                                             k_y\left({\cal L}_{\geq }[u]\right) \,.
\eeqn
\eps

\bpr
Pick any $\xi ,u$, and the corresponding $x(\cdot )$ and $y(\cdot )$, and any element
$\mu \in {\cal L}_{\leq }[u]$.
Let $t_k \rightarrow + \infty $, $\mu _k\rightarrow \mu $, with all $\mu _k\in \Inputset$, and
$u(t)\succeq \mu _k$ for all $t\geq t_k$.
By monotonicity of the system, for $t\geq t_k$ we have:
\be{firststep}
\begin{array}{rcl}
x(t,\xi ,u) &=& x(t - t_k,x(t_k,\xi ,u), u(\cdot + t_k) ) \\  &\succeq& 
x(t-t_k,x(t_k,\xi ,u),\mu _k) \,.
\end{array}
\ee 
In particular, if $x(s_\ell)\rightarrow \zeta $ for some sequence $s_\ell\rightarrow \infty $, it follows
that
$
\zeta \succeq \lim_{\ell\rightarrow \infty } x(s_\ell- t_k,x(t_k,\xi ,u),\mu _k)
=k_x(\mu _k)
$.
Next, taking limits as $k\rightarrow \infty $, and using continuity of $k_x$, this proves that
$\zeta \succeq k_x(\mu )$.
This property holds for every elements $\zeta \in  \Omega [x]$
and $\mu \in {\cal L}_{\leq }[u]$, so we have shown that
$k_x\left({\cal L}_{\leq }[u]\right) \preceq \Omega [x]$.
The remaining inequalities are all proved in an entirely analogous fashion.
\epr

\bp{asybehaviour}
Consider a monotone SISO system (\ref{cds}), with static I/S and I/O
characteristics $k_x(\cdot )$  and $k_y (\cdot )$.
Then, the I/S and I/O characteristics are nondecreasing, and
for each initial condition $\xi $ and each bounded input $u(\cdot )$,
the following holds:
\[
\begin{array}{rcl}
k_y( u_{\inf} ) &\leq& \liminf_{t \rightarrow + \infty} y (t, \xi ,u)  \\
&\leq&
\limsup_{t \rightarrow + \infty} y(t,\xi ,u) \leq k_y ( u_{\sup} ) \,.
\end{array}
\]
If, instead, outputs are ordered by $\geq$, then
the I/O static characteristic is nonincreasing, and
for each initial condition $\xi $ and each bounded input $u(\cdot )$,
the following inequality holds:
\[
\begin{array}{rcl}
k_y( u_{\sup} ) &\leq& \liminf_{t \rightarrow + \infty} y (t, \xi ,u)  \\
&\leq& \limsup_{t \rightarrow + \infty} y(t,\xi ,u) \leq k_y ( u_{\inf} ) \,.
\end{array}
\]
\eps

\bpr
The proof of the first statement is immediate from
Proposition~\ref{asybehaviour-general} and the properties:
$u_{\inf}\in {\cal L}_{\leq }[u]$, $u_{\sup}\in {\cal L}_{\geq }[u]$,
$\liminf_{t \rightarrow + \infty} y(t)\in \Omega [y]$, and
$\limsup_{t \rightarrow + \infty} y(t)\in \Omega [y]$, and 
the second statement is proved in a similar fashion.
\epr

\br{cauchy-rem}
It is an immediate consequence of Proposition~\ref{asybehaviour} that, if a
monotone system admits a static I/O characteristic $k$, and if there is a
class-$\ki$ function $\gamma $ such that $\abs{k(u)-k(v)}\leq \gamma (\abs{u-v})$ for all
$u,v$ (for instance, if $k$ is Lipschitz with constant $\rho $ one may pick as $\gamma $
the linear function $\gamma (r)=\rho r$) then the system has a Cauchy gain (in the
sense of~\cite{cauchy}) $\gamma $ on bounded inputs.
\er

\section{Feedback Interconnections}
\label{Feedback Interconnections of Monotone Systems}

\begin{figure}[ht]
\begin{center}
\setlength{\unitlength}{2500sp}%
\begin{picture}(2724,1866)(3889,-3715)
\put(4051,-2011){$u_1$}
\put(4051,-3661){$y_2$}
\put(6226,-2011){$y_1$}
\put(6226,-3661){$u_2$}
\thinlines
\put(4501,-2461){\framebox(1500,600){}}
\put(6001,-2161){\line( 1, 0){600}}
\put(6601,-2161){\line( 0,-1){1200}}
\put(6601,-3361){\vector(-1, 0){600}}
\put(4501,-3361){\line(-1, 0){600}}
\put(3901,-3361){\line( 0, 1){1200}}
\put(3901,-2161){\vector( 1, 0){600}}
\put(4501,-3661){\framebox(1500,600){}}
\put(5150,-2236){$\Sigma _1$}
\put(5150,-3436){$\Sigma _2$}
\end{picture}
\caption{Systems in feedback}
\label{interconnection}
\end{center}
\end{figure}
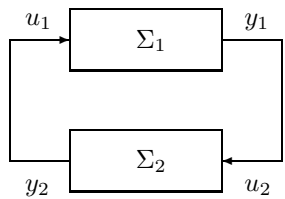

In this section, we study the stability of SISO monotone dynamical
systems connected in feedback
as in Fig.~\ref{interconnection}.
Observe that such interconnections need not be monotone.
Based on Proposition \ref{asybehaviour}, one of our main results will be the
formulation of a small-gain theorem for the feedback interconnection of a
system with monotonically increasing I/O static gain (positive path) and a
system with monotonically decreasing I/O gain (negative path).

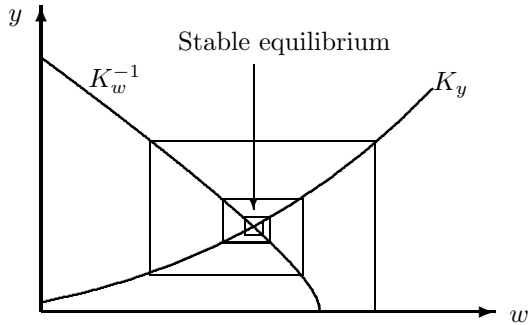
\begin{figure}[h,t]
\begin{center}
\setlength{\unitlength}{2000sp}%
\begin{picture}(6210,3947)(1396,-5596)
\thicklines
\curve(1801,-2311,4436,-4411,5250,-5461)
\curve(1800,-5350,4436,-4411,6650,-2700)
\put(1801,-5461){\vector( 1, 0){5625}}
\put(1801,-5461){\vector( 0, 1){3800}}
\thinlines
\put(4436,-2401){\vector( 0,-1){1800}}
\put(5941,-5461){\line( 0, 1){2115}}
\put(5941,-3346){\line(-1, 0){2790}}
\put(3151,-3346){\line( 0,-1){1665}}
\put(3151,-5011){\line( 1, 0){1890}}
\put(5041,-5011){\line( 0, 1){945}}
\put(5041,-4066){\line(-1, 0){990}}
\put(4051,-4066){\line( 0,-1){540}}
\put(4051,-4606){\line( 1, 0){585}}
\put(4636,-4606){\line( 0, 1){315}}
\put(4636,-4291){\line(-1, 0){315}}
\put(4321,-4291){\line( 0,-1){225}}
\put(4321,-4516){\line( 1, 0){225}}
\put(4546,-4516){\line( 0, 1){180}}
\put(3500,-2221){Stable equilibrium}
\put(1396,-1861){$y$}
\put(7606,-5596){$w$}
\put(2386,-2671){$K_w^{-1}$}
\put(6661,-2716){$K_y$}
\end{picture}
\caption{I/O characteristics in $(w,y)$ plane: negative
feedback} %; the graph shown for $K_z$ is that of the inverse map}
\label{lowgain}
\end{center}
\end{figure}

\bt{sgt3}
Consider the following interconnection of 
two SISO
dynamical systems
\be{mdsfeedback}
\begin{array}{rcl}
\dot {x} &=& f_x ( x, w )\,,\quad y=h_x(x) \\
\dot {z} &=& f_z ( z, y ) \,,\quad \; w=h_z(z)
\end{array}
\ee
with $\Inputset_x=\Outputset_z$ and
$\Inputset_z=\Outputset_x$.
Suppose that:
\ben
\item
the first system is monotone when its input $w$ as well as output
$y$ 
are ordered according to the ``standard order'' induced by the positive real semi-axis;
\item
the
second system is monotone when its input $y$ is ordered according to the standard order induced by the
positive real semi-axis 
and its output $w$ is ordered by the opposite order, viz. the one induced by the negative real semi-axis;
\item
the respective static I/S characteristics $k_x (\cdot )$ and $k_z(\cdot )$ exist
(thus, the static I/O characteristics $k_y (\cdot )$ and $k_w(\cdot )$ exist too and
are respectively monotonically increasing and monotonically decreasing);
and
\item
every solution of the closed-loop system is bounded.
\een
Then, system~(\ref{mdsfeedback}) has a globally attractive
equilibrium provided that the following scalar discrete time dynamical system, 
evolving in $\Inputset_x$:
\be{eqcond}
u_{k+1} = \left(k_w \circ k_y \right)( u_k )  
\ee
has a unique globally attractive equilibrium $\bar{u}$. 
\ets
For a graphical interpretation of condition (\ref{eqcond}) see
Fig.~\ref{lowgain}.

\bpr
Equilibria of (\ref{mdsfeedback}) are in one to one 
correspondence with solutions of $k_w(k_y(u)) = u$, viz. equilibria of
(\ref{eqcond}). Thus, existence and uniqueness of the equilibrium follows
from the GAS assumption on (\ref{eqcond}).

We need to show that such an equilibrium is globally attractive. Let $\xi \in
\R^{n_x} \times \R^{n_z}$ be an arbitrary initial condition and let $y_+ :=
\limsup_{t \rightarrow + \infty} y (t,\xi )$ and 
$y_- := \liminf_{t \rightarrow + \infty} y (t,\xi )$.
Then,
$w_+:=\limsup_{t\rightarrow +\infty} w (t,\xi )$ and 
$w_- := \liminf_{t \rightarrow + \infty} w(t,\xi )$
satisfy by virtue of the second part of Proposition \ref{asybehaviour},
applied to the $z$-subsystem:
\be{tobequoted}
k_w ( y_+ ) \leq w_-  \leq  w_+  \leq k_w ( y_- ).
\ee
 An analogous argument, applied to the 
$x$-subsystem, yields: $k_y (w_-) \leq y_- \leq y_+ \leq k_y (w_+)$ and by combining
this with the inequalities for $w_+$ and $w_-$we end up with:
\[
k_y( k_w (y_+)) \leq y_- \leq y_+ \leq k_y (k_w ( y_- )) \,.
\]  
By induction we have, after an even number $2n$ of iterations of the above
argument:
\[
( k_y \circ k_w )^{2n} (y_-) \leq y_- \leq y_+ \leq (k_y \circ k_w)^{2n} (y_+)\,.
\]
By letting $n \rightarrow + \infty$ and exploiting global attractivity of (\ref{eqcond})
we have $y_- = y_+$. Equation (\ref{tobequoted}) yields $w_- = w_+$. 
Thus there exists $\bar{u}$, such that:
\be{limitsexist}
\begin{array}{rcl}
\bar{u} &=& \lim_{t \rightarrow + \infty} y(t,\xi ) \\
 k_w (\bar{u}) &=& \lim_{t \rightarrow + \infty} w(t,\xi )\,.
\end{array}
\ee 
Let $z_e$ be the (globally asymptotically stable) equilibrium (for the
$z$-subsystem) corresponding to the constant input $y(t) \equiv \bar{u}$ and
$x_e$ the equilibrium for the $x$-subsystem relative to the input $w(t)
\equiv k_w (\bar{u})$.
Clearly $\eta :=[x_e , z_e ]$ is the unique equilibrium of (\ref{mdsfeedback}).
The fact that $[ x(t,\xi ), z(t,\xi ) ] \rightarrow \eta $ now follows from
Proposition~\ref{cics}. 
\epr

\br{boundassumption} We remark that traditional small-gain theorems
also provide sufficient conditions for global existence and boundedness
of solutions. In this respect it is of interest to notice that, for monotone
systems, boundedness of trajectories follows at once provided that at least
one of the interconnected systems has a uniformly bounded output map  
(this is always the case for instance if the state space of the corresponding system
is compact). However, when both output maps are unbounded,
boundedness of trajectories needs to be proved with different techniques. 
The following Proposition addresses this issue and provides additional conditions
which together with the small-gain condition allow to conclude boundedness of trajectories.
\er

We say that the I/S characteristic $k_x(\cdot )$ is \emph{unbounded} (relative to
$X$) if for all $\xi \in X$ there exist $u_1, u_2 \in \Inputset$ so that 
$k_x(u_1)\succeq\xi \succeq k_x(u_2)$.  

\bl{ISSproperty}
Suppose that the system~(\ref{cds}) is endowed with an unbounded I/S static
characteristic and that inputs are scalar ($\Bu=\R$ with the usual order).
Then, for any 
$\xi \in X$ there exists $\bar{\xi } \in X$ so that for any input $u$:
\be{isslike}
\phi (t,\xi ,u) \preceq \max
\{ \bar{\xi }, k_x \left( {\textstyle{\sup_{\tau \in [0,t]}}} u (\tau )\right) \}
\;\;\forall\,t\geq 0\,.
\ee
An analogous property holds with $\preceq$ replaced by $\succeq$ and $\sup$'s
replaced by $\inf$'s. 
\els

\bpr
Let $\xi \in X$ be arbitrary. As $k_x$ is unbounded there exists $\bar{u}$ such
that $\xi \preceq k_x (\bar{u} ):= \bar{\xi } $. 
Pick any input $u$ and any $t_0 \geq 0$, and
let $\mu :=\sup_{\tau \in [0,t_0]} u(\tau )$.
There are two possibilities: 
$\mu  \leq \bar{u}$ or $\mu  \geq \bar{u}$.
By monotonicity with respect to initial conditions and inputs,  the first case
yields: 
\be{firstcase}
\phi (t_0,\xi ,u) \preceq \phi (t_0,\bar{\xi }, \bar{u} ) = \bar{\xi }  \,.
\ee
So we assume from now on that $\mu  \geq \bar{u}$.
We introduce the input $U$ defined as follows:
$U(t):=\mu $ for all $t\leq t_0$, and $U(t)=u(t)$ for $t>t_0$.
Notice that $U\succeq u$, and also that
$\phi (t_0,k_x(\mu ),U)=k_x(\mu )$, because the state $k_x(\mu )$ is by definition an
equilibrium of $\dot x=f(x.\mu )$ and $U(t)\equiv \mu $ on the interval $[0,t_0]$.
We conclude that
\be{seccase}
\phi (t_0,\xi ,u) \preceq \phi (t_0, k_x(\bar{u}),U))= k_x (\mu )
\ee
and~(\ref{isslike}) follows combining (\ref{firstcase}) and (\ref{seccase}).
The statement for $\succeq$ is proved in the same manner.
\epr

\bp{boundtheorem} 
Consider the feedback interconnection of two SISO monotone dynamical
systems as in (\ref{mdsfeedback}), and assume that the orders in both
state-spaces are bounded.
Assume that the systems are endowed with \emph{unbounded} 
I/S static characteristics $k_x(\cdot )$ and $k_z (\cdot )$ respectively. 
If the small gain condition of Theorem \ref{sgt3} is satisfied then solutions
exist for all positive times, and are bounded.
\eps

Clearly, the above result allows to apply Theorem \ref{sgt3} also to classes
of monotone systems 
for which boundedness of trajectories is not a priori known.

\bpr
We show first that solutions are upper-bounded. A symmetric argument can be used for 
determining a lower bound. Let $\xi , \zeta $ be arbitrary initial conditions for 
the $x$ and $z$ subsystems. Correspondingly solutions are maximally defined over some
interval $[0,T)$. Let $t$ be arbitrary in $[0,T)$. By Lemma~\ref{ISSproperty},
equation (\ref{isslike}) holds, for each of the systems.
Moreover, composing (\ref{isslike}) (and its counter-part for lower-bounds) 
with the output map yields, for suitable constants
$\bar{y}, \bar{w}, \underline{y}, \underline{w}$ which only depend upon
$\xi ,\zeta $. 
\begin{eqnarray}
\label{aeq}
y(t,\xi ,w) & \leq &  \max \leftn\{ \bar{y}, k_y \leftn( 
{\textstyle{\max_{\tau \in [0,t]}}} w(\tau ) \rightn) \rightn\} \\
\label{beq}
w(t,\zeta , y ) & \leq & \max \leftn\{ \bar{w}, k_w \leftn( 
{\textstyle{\min_{\tau \in [0,t]}}} y(\tau ) \rightn) \rightn\} \\
\label{ceq}
y(t,\xi ,w) & \geq & \min \leftn\{ \underline{y}, k_y \leftn( 
{\textstyle{\min_{\tau \in [0,t]}}} w(\tau ) \rightn) \rightn\} \\
\label{deq}
w(t,\zeta , y ) & \geq & \min \leftn\{ \underline{w}, k_w \leftn( 
{\textstyle{\max_{\tau \in [0,t]}}} y(\tau ) \rightn) \rightn\}.
\end{eqnarray}
Substituting equation (\ref{beq}) into (\ref{aeq}) gives:
\be{firstsub}
y(t,\xi ,w) \leq \max \leftn\{ \bar{y}, k_y (\bar{w} ), 
k_y \circ k_w \leftn( 
{\textstyle{\min_{\tau \in [0,t]}}}y(\tau ) \rightn) \rightn\}
\ee
and substitution of (\ref{ceq}) into (\ref{firstsub}) yields
(using that $k_y \circ k_w$ is a nonincreasing function):
\be{secsub}
\begin{array}{rcl}
y(t,\xi ,w) &\leq& \max \Big \{ \bar{y}, k_y (\bar{w} ), k_y \circ k_w ( \underline{y} ) , \\ & &
k_y \circ k_w \circ k_y ( \min_{\tau \in [0,t]} w(\tau ) ) \Big \}.
\end{array}
\ee
Finally equation (\ref{deq}) into (\ref{secsub}) yields:
\be{thirdsub}
y(t,\xi ,w) \leq \max \left\{ a,
\rho \circ \rho \,
\left( \max_{\tau \in [0,t]} y(\tau ) \right) \right\}\,,
\ee
where we are denoting $\rho :=k_y \circ k_w$
and
\[
a
:=\max\left\{\bar{y},k_y(\bar{w}),k_y\circ k_w(\underline{y}),k_y\circ k_w\circ k_y(\underline{w})\right\} \,.
\]
Let $y_{e}$ be the output value of the $x$-subsystem, corresponding to the
unique equilibrium of the feedback interconnection (\ref{mdsfeedback}). 

Notice that attractivity of  (\ref{eqcond}) implies attractivity of
$y(t+1) = k_y \circ k_w (y(t)) := \rho (y(t))$ and a fortriori of 
\be{reducedsystem}
 y(t+1) = \rho \circ \rho ( y(t) ).
\ee 
We claim that
$      y > y_e  \Rightarrow  \rho \circ \rho ( y ) < y$.
By attractivity, there exists some $y_1 > y_e$ such that
 $\rho \circ \rho ( y_1 ) - y_1 < 0$ (otherwise all trajectories
of (\ref{reducedsystem}) starting from $y > y_e$ would be monotonically
increasing, which is absurd). 
Now, assume by contradiction that there exists also some $y_2>y_e$ such that
$\rho \circ \rho (y_2)-y_2>0$.  Then, as $\rho $ is a continuous function, there would exist
an $y_0\in (y_1,y_2)$ (or in $(y_2,y_1)$ if $y_2<y_1$) such that
$\rho \circ \rho (y_0)=y_0$.  This clearly violates attractivity (at $y_e$) of
(\ref{reducedsystem}), since $y_0$ is an equilibrium point. 
So the claim is proved.

Let $M:=\max_{\tau \in [0,t]} y(\tau )$, so $M=y(\tau _0)$ for some $\tau _0\in [0,t]$.
Therefore~(\ref{thirdsub}) at $t=\tau _0$ says that
$y(\tau _0)\leq \max\{a,\rho \circ \rho (y(\tau _0))\}$, and the previous claim applied at
$y=y(\tau _0)$ gives that $y(\tau _0) \leq \max \{ a, y_e \} $ (by
considering separately the cases $y (\tau _0) > y_e$ and $y(\tau _0) \leq
y_e$).
As $y(t)\leq y (\tau _0)$, we conclude that $y(t)\leq \max \{ a, y_e \}$.
This shows that $y$ is upper bounded by a function which depends only on the
initial states of the closed-loop system.
Analogous arguments can be used in order to show that $y$ is lower bounded,
and by symmetry the same applies to $w$. 
Thus, over the interval $[0,T)$ the $x$ and $z$ subsystems are fed by bounded
inputs and by monotonicity (together with the existence of I/S static
characteristics) this implies, 
by Proposition~\ref{bibs},
that $T= + \infty$ and that trajectories are uniformly bounded.  
\epr

\section{An Application}
\label{An Application}

A large variety of eukaryotic cell signal transduction processes
employ ``Mitogen-activated protein kinase (MAPK) cascades,''
which play a role in some of the most fundamental processes of life
(cell proliferation and growth, responses to hormones, etc).
A MAPK cascade is a cascade connection of three SISO systems, each of which
is (after restricting to stoichiometrically conserved subsets) either a one- or
a two-dimensional system, see~\cite{ferrell,kholodenko}.
We will show here that the two-dimensional case gives rise to monotone systems
which admit static I/O characteristics.
(The same holds for the much easier one-dimensional case, as follows from the
results in~\cite{cauchy}.)

After nondimensionalization, the basic system to be studied is a system
as in~(\ref{nlsys}), where the functions $\theta _i$ are of the type
$
\theta _i(r)=\frac{a_i r}{1+b_i r}
$,
for various positive constants $a_i$ and $b_i$.
It follows from Proposition~\ref{mapk-basic-piece-is-monotone} that 
our systems (with output $y$)
are monotone, and therefore {\em every MAPK cascade is monotone}.

We claim, further, that each such system has a static I/O characteristic.
(The proof that we give is based on a result that is specific to
two-dimensional systems; an alternative argument, based upon a triangular
change of variables as mentioned earlier, would also apply to more arbitrary
signaling cascades, see~\cite{part2}.)
It will follow, by basic properties of cascades of stable systems, that
the cascades have the same property.  Thus, the complete theory developed in
this paper, including small gain theorems, can be applied to MAPK cascades.

\bp{main-prop-mapk}
For any system of the type~(\ref{nlsys}), and each constant input $u$, there
exists a unique globally asymptotically stable equilibrium inside $\Delta $.
\eps

\bpr
As the set $\Delta $ is positively invariant, the Brower Fixed-Point
Theorem ensures existence of an equilibrium.
We next consider the Jacobian $\differential f$ of
$f$. 
It turns out that for all $(x^1,x^2) \in \Delta $ and all  $u \geq 0$ 
\begin{eqnarray}
\textrm{tr} ( \differential f) &=& - u D\theta _1 (x^1) - D \theta _2 (1-x^1-x^2) + \nonumber \\ &-& D \theta 
_4(x^2) -u D \theta _3 (1-x^1-x^2)  
< 0, \quad \nonumber \\
\textrm{det} ( \differential f) &=&  u^2 D\theta 
_1(x^1)
D \theta _3 (1-x^1-x^2) + \nonumber \\ &+& u D \theta _1 (x^1) D \theta _4 (x^2) + \nonumber \\ &+& D \theta _2 (1-x^1-x^2) D \theta _4 (x^2)  > 0 . \nonumber
\end{eqnarray}
The functions $\theta _i$ are only defined on intervals of the form
$(-1/b_i,+\infty )$.
However, we may assume without loss of generality that they are each defined
on all of $\R$, and moreover that their derivatives are positive on all of
$\R$.  Indeed, let us pick any continuously differentiable functions
$\sigma _i:\R\rightarrow \R$, $i=1,2,3,4$ with the properties that $\sigma _i'(p)>0$ 
for all $p\in \R$, $\sigma _i(p)=p$ for all $p\geq 0$, and the image of $\sigma _i$ is
contained in $(-1/b_i,+\infty )$.  Then we replace each $\theta _i$ by the composition
$\theta _i\circ \sigma _i$.  

Note that the functions $\theta _i\circ \sigma _i$ have an everywhere positive derivative,
so $\textrm{tr} ( \differential f)$ and $\textrm{det} ( \differential f)$ are
everywhere negative and positive, respectively, in $\R^2$.
So $\differential f$ is Hurwitz everywhere.
The Markus-Yamabe conjecture on global asymptotic stability (1960) was that
if a ${\cal C}^1$ map $\R^n \rightarrow  \R^n$ has a zero at a point $p$, and its Jacobian
is everywhere a Hurwitz matrix, then $p$ is a globally asymptotically stable
point for the system $\dot x=f(x)$.
This conjecture is known to be false in general, but true in dimension two,
in which case it was proved simultaneously by Fessler, Gutierres, and Glutsyuk
in 1993, see e.g.~\cite{fessler}.
Thus, our (modified) system has its equilibrium as a globally asymptotically
stable attractor in $\R^2$.
As inside the triangle $\Delta $, the original $\theta _i$'s coincide with
the modified ones, this proves global stability of the original system
(and, necessarily, uniqueness of the equilibrium as well).
\epr

\begin{figure}[ht]
\begin{center}
\vskip1cm
\includegraphics[width=6cm,height=3cm]{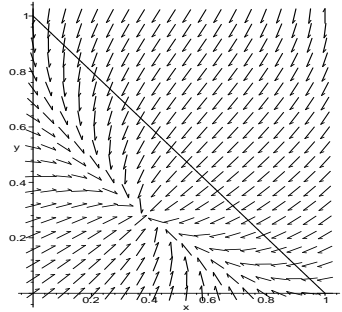}
\caption{Direction field for example}
\label{field-mapk-3}
\end{center}
\end{figure}

As an example,
Fig.~\ref{field-mapk-3} shows the phase plane of the system (the diagonal line
indicates the boundary of the triangular region of interest), when
coefficients have been chosen so that the equations are:
$
\dot x_1=- 1.0\,{\frac {x_1}{1+x_1}}+2\,{\frac {1-x_1-x_2}{3-x_1-x_2}}$
and 
$\dot x_2={\frac {1-x_1-x_2}{2-x_1-x_2}}-2\,{\frac {x_2}{2+x_2}}$.

As a concrete illustration, 
let us consider the open-loop system with these equations:
\beqn
\dot x_1 &=& {\frac{\coefV_2\,(100-x_1)}{\coefK_2+(100-x_1)}}
-\frac{\coefG_1 x_1}{\coefK_1+x_1}\frac{\coefG_2+u}{\coefG_4+u}\\
\dot y_1 &=& {\frac{\coefV_6\,(300-y_1-y_3)}{\coefK_6+(300-y_1-y_3)}}-{\frac{\coefk_3\,(100-x_1)\,y_1}{\coefK_3+y_1}}\\
\dot y_3 &=& {\frac{\coefk_4\,(100-x_1)\,(300-y_1-y_3)}{\coefK_4+(300-y_1-y_3)}}-{\frac{\coefV_5\,y_3}{\coefK_5+y_3}}\\
\dot z_1 &=& {\frac{\coefV_{10}\,(300-z_1-z_3)}{\coefK_{10}+(300-z_1-z_3)}}-{\frac{\coefk_7\,y_3\,z_1}{\coefK_7+z_1}}\\
\dot z_3 &=&  {\frac{\coefk_8\,y_3\,(300-z_1-z_3)}{\coefK_8+(300-z_1-z_3)}}-{\frac{\coefV_9\,z_3}{\coefK_9+z_3}}.
\eeqn
This is the model studied in~\cite{kholodenko}, from which we also borrow the
values of constants (with a couple of exceptions, see below):
$\coefG_1=0.22$,
$\coefG_2=45$,
$\coefG_4=50$,
$\coefK_1=10$,
$\coefV_2=0.25$,
$\coefK_2=8$,
$\coefk_3=0.025$,
$\coefK_3=15$,
$\coefk_4=0.025$
$\coefK_4=15$,
$\coefV_5=0.75$,
$\coefK_5=15$,
$\coefV_6=0.75$,
$\coefK_6=15$,
$\coefk_7=0.025$,
$\coefK_7=15$,
$\coefk_8=0.025$,
$\coefK_8=15$,
$\coefV_9=0.5$,
$\coefK_9=15$,
$\coefV_{10}=0.5$,
$\coefK_{10}=15$.
Units are as follows: concentrations and Michaelis constants ($\coefK$'s) are
expressed in nM, catalytic rate constants ($\coefk$'s) in $s^{-1}$, and
maximal enzyme rates ($\coefV$'s) in $nM.s^{-1}$.
The paper~\cite{kholodenko} showed that oscillations may arise in this system
for appropriate values of negative feedback gains.
(We have slightly changed the input term, using coefficients $\coefG_1$,
$\coefG_2$, $\coefG_4$, because we wish to emphasize the open-loop system
before considering the effect of negative feedback.)

\begin{figure}[ht]
\begin{center}
\includegraphics[width=6cm,height=6cm]{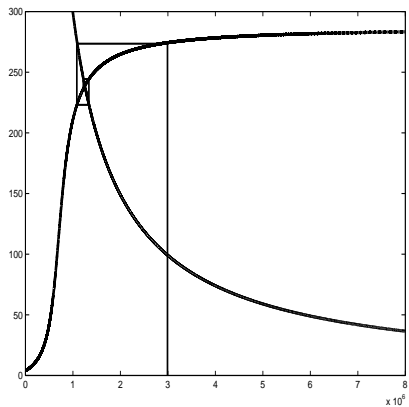}
\caption{I/O characteristic and small-gain for MAPK example}
\label{sgt-mapk}
\end{center}
\end{figure}

\begin{figure}[ht]
\begin{center}
\vskip1cm
\includegraphics[width=6cm,height=6cm]{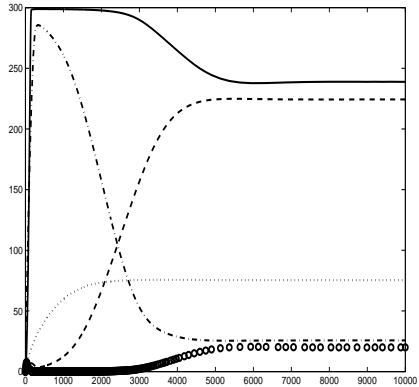}
\caption{Simulation of MAPK system under negative feedback satisfying
small-gain conditions.
Key:
$x_1$ dots,
$y_1$ dashes,
$y_2$ dash-dot,
$z_1$ circles,
$z_3$ solid}
\label{simul-mapk}
\end{center}
\end{figure}

Since the system is a cascade of elementary MAPK subsystems, we know that our
small-gain result may be applied.  Figure~\ref{sgt-mapk}
shows the I/O characteristic of this system, as well as the characteristic
corresponding to a feedback $u=\frac{K}{1+y}$, with the gain $K=30000$.
It is evident from this planar plot that the small-gain condition is satisfied
- a ``spiderweb'' diagram shows convergence.
Our theorem then guarantees global attraction to a unique equilibrium.
Indeed, Figure~\ref{simul-mapk} shows a typical state trajectory.

\section{Relations to Positivity}
\label{Relations to Positivity}

In this section we investigate the relationship between the notions of 
cooperative and positive systems.
Positive linear systems (in continuous as well as discrete time) have
attracted much attention in the control literature, see for instance
\cite{aeyels,farina-rinaldi-book,luenberger,muratori-rinaldi,piccardi-rinaldi,valcher}.
We will say that a finite dimensional linear system, possibly time-varying,
\be{linsys}
\dot {x} = A(t) x + B (t) u
\ee
(where the entries of the $n\times  n$ matrix $A$ and the $n\times  m$ matrix $B$ are
Lebesgue measurable locally essentially bounded functions of time)
is \emph{positive} if the positive orthant is forward invariant for 
positive input signals;
in other words, for any $\xi \succeq 0$ and any $u(t) \succeq 0$ 
($\succeq$ denotes here the partial 
orders induced by the positive orthants),
and any $t_0
\in \R$ it holds that $\phi (t,t_0, \xi ,u) \succeq 0$ for all $t \geq t_0$.

Let say that~(\ref{linsys}) is a {\em Metzler system} if
$A(t)$ is a Metzler matrix, i.e., $A_{ij}(t) \geq 0 $ for all $i \neq j$, and
$B_{ij}(t)\geq 0$ for all $i,j$, for almost all $t\geq 0$.
It is well known for time-invariant systems ($A$ and $B$ constant),
see for instance~\cite{luenberger}, Chapter 6, or \cite{aeyels} for a
recent reference, that a system is positive if and only if it is
a Metzler system.  This also holds for the general
case, and we provide the proof here for completeness.
For simplicity in the proof, and because we only need this case, we make a
continuity assumption in one of the implications.

\bl{Metzler-lemma}
If~(\ref{linsys}) is a Metzler system then it is positive.
Conversely, if~(\ref{linsys}) is positive and $A(\cdot )$ and $B(\cdot )$ are
continuous, then~(\ref{linsys}) is a Metzler system. 
\els

\bpr
Let us prove sufficiency first.
Consider first any trajectory $x(\cdot )$ with $x(s) \gg 0$, any fixed $T>s$, and
any input $u(\cdot )$ so that $u(t) \geq 0$ for all $t \geq s$.
We need to prove that $x(T) \succeq 0$.
Since $A(t)$ is essentially bounded (over any bounded time-interval) and
Metzler, there is an $r>0$ 
such that $rI + A(t) \geq 0$ for almost all $t \in [s,T]$, where ``$\geq $'' is meant
elementwise. 
Consider $z(t):= \textrm{exp}(r(t-s)) x(t)$ and 
$v(t):= \textrm{exp}(r(t-s))u(t)$,
and note that $z(s) = x(s) \gg 0$ and $v(t) \geq 0$ for all $t\geq s$.
We claim that $z(t) \succeq 0$ for all $t \in [s,T]$.
Let $\tau >s$ be the infimum of the set of $t$'s such that $z(t) \nsucceq 0$ and
assume, by contradiction, $\tau < +\infty$.
By continuity of trajectories, $z(\tau ) \succeq 0$. 
Moreover 
$z(\tau )=z(s)+\int_s^\tau \dot z(t) dt 
= z(s) + \int_s^\tau (rI+A(t))z(t) + B(t)v(t) \, dt \succeq z(s) \gg 0 $,
and therefore there exists an interval $[\tau ,\tau + \varepsilon ]$
such that $z(t) \succeq  0$ for all $t \in [\tau ,\tau +\varepsilon ]$.
But this is a contradiction, unless $\tau =+\infty$ as claimed.
By continuous dependence with respect to initial conditions, and closedness of
the positive orthant, the result carries over to any initial condition
$x(s)\succeq 0$. 
For the converse implication, denote with $\Phi (t,s)$ the fundamental solution
associated to $A(t)$ ($\partial \Phi / \partial t = A(t) \Phi $, $\Phi (s,s)=\textrm{I}$).
Using $u \equiv 0$ we know that $\Phi (t,s) \geq 0$ whenever $t \geq s$ (``$\geq$'' is
meant here elementwise).
Therefore also $[\Phi (t,s)-I]_{ij} \geq 0$ for all $i \neq j$.
Since 
$A(\tau ) = \left . \fract{\partial}{ \partial t } \right |_{t=\tau } \Phi (t,\tau )  = \lim_{t \rightarrow 0}
\frac{\Phi (t,\tau )-I}{t}$
for all $\tau $, it follows that $A(\tau )_{ij} \geq 0$ for all $i\neq j$. 
Consider a solution with $x(s)=0$, $u$ constant $\geq 0$, for $t \geq s$.
Since $x(t) \succeq 0$, also $(1/(t-s))x(t)\succeq 0$, and therefore, taking
limits as $t\searrow0$, $\dot {x}(s) \succeq 0$ (the derivative exists by the
continuity assumption).
But $\dot {x}(s) = A(s) x(s) + B(s)u$, and $x(s)=0$, so $B(s)u \succeq 0$ for all
such $u$, i.e. $B(s) \succeq 0$.
\epr

Thus, by virtue of Theorem \ref{infinitesimal-derivs}, a time-invariant linear
system is cooperative if and only if it is positive. 
The next result is a system-theoretic analog of the fact that a differentiable
scalar real function is monotonically increasing if and only if its derivative
is always nonnegative.

We say that a system~(\ref{cds}) is {\em incrementally positive} (or
``variationally positive'') if, for every
solution $x(t)=\phi (t,\xi ,u)$ of~(\ref{cds}), the linearized system
\be{variational}
\dot z\;=\;A(t)z + B(t)v
\ee
where $A(t)=\frac{ \partial f }{\partial x} (x(t),u(t))$ and
$B(t)=\frac{ \partial f}{\partial u} (x(t),u(t))$,
is a positive system.
\bp{moniffpos}
Suppose that $\Bu=\R^m$, $\Inputset$ satisfies an approximability property,
and that both $\Xo$ and $\Uo=\inter \Inputset$ are order-convex. Let $f(x,u)$
be continuously differentiable.
Then system~(\ref{cds}) is cooperative if and only if
it is incrementally positive.
\eps

\bpr 
Under the given hypotheses, a system is cooperative iff $\frac{\partial f}{\partial x}
(x,u)$ is a Metzler matrix, and every entry of $\frac{\partial f}{\partial u}(x,u)$ is
nonnegative, for all $x \in X$ and all $u \in \Inputset$, cf.\
Proposition~\ref{infinitesimal-derivs}.
Therefore, by the 
criterion for positivity of linear time-varying systems, this implies 
that (\ref{variational}) is a positive linear time-varying system along
any trajectory of (\ref{cds}). 

Conversely, pick an arbitrary $\xi $ in $X$  and any input of the form
$u(\cdot )=\bar{u} \in \Inputset$.
Suppose that~(\ref{variational}) is a positive linear time-varying
system along the trajectory $x(t)=x(t,\xi ,u)$ (this system has continuous
matrices  $A$ and $B$ because $u$ is constant). 
Then, by the positivity criterion of linear time-varying systems, 
for all $t \geq 0$ we have $\frac{\partial f}{\partial x}(x(t),\ubar)$ is Metzler and
$\frac{\partial f}{\partial u}(x(t),\ubar)\succeq 0$. 
Finally, evaluating the Jacobian at $t=0$ yields that
$\frac{\partial f}{\partial x}(\xi ,\ubar)$ is Metzler and
$\frac{\partial f}{\partial u}(\xi ,\ubar)$ is nonnegative.
Since $\xi $ and $\ubar$ were arbitrary, we have the condition for cooperativity
given in Proposition \ref{infinitesimal-derivs}.
\epr

\br{alternative}
Looking at cooperativity as a notion of ``incremental positivity'' one can
provide an alternative proof of the infinitesimal condition for cooperativity,
based on the positivity of the variational equation.
Indeed, assume that each system~(\ref{variational}) is a positive linear
time-varying system, along trajectories of (\ref{cds}).
Pick arbitrary initial conditions $\xi _1 \succeq \xi _2 \in X$ and inputs
$u_1\geq u_2$.   Let 
$\Phi (h) := \phi (t,\xi _2 + h ( \xi _1 - \xi _2 ) ,u_2+ h (u_1 - u_2 ) )$.
We have (see e.g.\ Theorem 1 in \cite{mct}) that
$\phi (t,\xi _1,u_1) - \phi (t,\xi _2 , u_2 ) = \Phi (1 ) - \Phi (0)
= \int_0^1 \Phi ^\prime (h) \, dh
= \int_0^1 z_h (t, \xi _1-\xi _2, u_1 - u_2 ) \, dh 
$,
where $z_h$ denotes the solution of (\ref{variational}) when $\frac{\partial f}{\partial u}(x,u) $ and $\frac{\partial f}{\partial u}(x,u)$ are evaluated 
along $\phi (t,\xi _2 + h ( \xi _1 - \xi _2 ), u_2 + h (u_1 - u_2 ) )$.
Therefore, by positivity, 
and monotonicity of the integral, we have
$\phi (t,\xi _1,u_1 ) - \phi (t,\xi _2,u_2) \succeq 0$, as claimed.
\er   

We remark that monotonicity with respect to other orthants corresponds to
generalized positivity properties for linearizations, as should be clear by
Corollary~\ref{infinitesimal-derivs-general-orthant}.

\section*{Appendix: A Lemma on Invariance}
\setcounter{section}{1}
\renewcommand{\thesection}{\Alph{section}}

We present here a characterization of invariance of relatively closed sets,
under differential inclusions. T
he result is a simple adaptation of a well-known condition, and is expressed
in terms of appropriate tangent cones. 
We let $\Xo$ be an open subset of some Euclidean space $\R^n$
and consider set-valued mappings $F$ defined on $\Xo$:
these are mappings which assign some subset $F(x)\subseteq \R^n$ to each $x\in \Xo$.
Associated to such mappings $F$ are {\em differential inclusions}
\be{diff-incl}
\dot x \in  F(x)
\ee
and one says that a function $x:[0,T]\rightarrow \Xo$ is a {\em solution}
of~(\ref{diff-incl}) if $x$ is an absolutely continuous function with the
property that $\dot x(t)\in  F(x(t))$ for almost all $t\in [0,T]$.
A set-valued mapping $F$ is {\em compact-valued} if $F(\xi )$ is a compact set,
for each $\xi \in  \Xo$, and 
it is {\em locally Lipschitz} if the
following property holds: for each 
compact subset $C\subseteq \Xo$
there is some constant $k$ such that
$F(\xi ) \subseteq   F(\zeta )+k\abs{\xi -\zeta }B$
for all $\xi ,\zeta \in  C$,
where $B$ denotes the unit ball in $\R^n$.
(We use $\abs{x}$ to denote Euclidean norm in $\R^n$.)
Note that when $F(x)=\{f(x)\}$ is single-valued, this is the usual definition
of a locally Lipschitz function.
More generally, suppose that $f(x,u)$ is locally Lipschitz in
$x\in \Xo$, locally uniformly on $u$, and pick any compact subset $D$ of the
input set $\Inputset$; then $F_D(x)=\{f(x,u), u\in  D\}$ is locally Lipschitz
and compact-valued.
We say that the set-valued mapping $F$ defined on $\Xo$ is
{\em locally bounded} if for each compact subset $C\subseteq \Xo$
there is some constant $k$ such that
$F(\xi )\subseteq  k B$ for all $\xi  \in  C$.
When $F$ has the form $F_D$ as above, it is locally bounded, since
$F_D(\xi )\subseteq  f(C\times  D)$, and, $f$ being continuous, the latter set is compact.

Let $\So$ be a (nonempty) closed subset relative to $\Xo$, that is,
$\So=S \bigcap \Xo$ for some closed subset $S$ of $\R^n$.
We wish to characterize the property that solutions which start in the set $\So$
must remain there.
Recall that
the subset $\So$ is said to be {\em strongly invariant} under the differential
inclusion~(\ref{diff-incl}) if the following property holds:
for every solution $x:[0,T]\rightarrow \Xo$ which has the property that $x(0)\in  \So$, it
must be the case that $x(t)\in \So$ for all $t\in [0,T]$.

Note that a vector $v$ belongs to $\tanSx$ 
(the ``Bouligand'' or ``contingent'' tangent cone)
if and only if there is a sequence
of elements $v_i\in \Xo$, $v_i\rightarrow v$ and a sequence $t_i\searrow0$ such that
$\xi +t_i v_i\in  \So$ for all $i$.
Further, $\tanSx=\R^n$ when $x$ is in the interior of $\So$ relative to $\Xo$
(so only boundary points are of interest).

\bt{main-theorem-invariance}
Suppose that $F$ is a locally Lipschitz, compact-valued, and locally bounded
set-valued mapping on the open subset $\Xo\subseteq \R^n$, and $\So$ is a closed
subset of $\Xo$. 
Then, the following two properties are equivalent:
\ben
\item
$\So$ is strongly invariant under $F$.
\item
$F(\xi )\subseteq  \tanSx$ for every $\xi \in  \So$.
\een
\ets

Just for purposes of the proof, let us say that a set-valued mapping $F$
is ``nice'' if $F$ is defined on all of $\R^n$ and it satisfies the following
properties: $F$ is locally Lipschitz, compact-valued, convex-valued,
and globally bounded ($F(\xi )\subseteq  k B$ for all $\xi \in \R^n$, for some $k$).
Theorem 4.3.8 in \cite{clsw} establishes that Properties 1 and 2 in the
statement of Theorem~\ref{main-theorem-invariance} are equivalent, and are
also equivalent to:
\be{co-tan}
F(\xi ) \subseteq  \co \tanSx \quad \mbox{for \ every} \; \xi \in  \So
\ee
(``co'' indicates closed convex hull)
provided that $\So$ is a closed subset of $\R^n$ and $F$ is nice
(a weaker linear growth condition can be replaced for global boundedness,
c.f.\ the ``standing hypotheses'' in Section 4.1.2 of \cite{clsw}). 
We will reduce to this case using the following observation.

\bl{main-reduction-invariance}
Suppose that $F$ is a locally Lipschitz, compact-valued, and locally bounded
set-valued mapping on the open subset $\Xo\subseteq \R^n$, and $\So$ is a closed
subset of $\Xo$.
Let $M$ be any given compact subset of $\Xo$.
Then, there exist a nice set-valued $\hatF$ and a closed subset $\Sop$ of
$\R^n$ such that the following properties hold:
\be{F-included-hatF-on-M}
F(\xi ) \subseteq  \hatF(\xi ) \quad\quad \forall \, \xi \in  M
\ee
\be{S-vs-S'}
M\bigcap  \So \,\subseteq \, \Sop \,\subseteq \, \So
\ee
\be{possible-tangents}
\forall\,\xi \in  \Sop,
\,\mbox{either}\;
\hatF(\xi )=\{0\}
\;\mbox{or}\;
\tanSx = \tanSpx
\ee
\be{possible-tangents-M}
\tanSx = \tanSpx\quad\quad\forall\,\xi \in  M\,.
\ee
and
$\So$ strongly invariant under $F$
implies
$\Sop$ strongly invariant under $\hatF$.
\els

\bpr
Consider the convexification $\widetilde F$ of $F$; this is the set-valued function on
$\Xo$ which is obtained by taking the convex hull of the sets $F(\xi )$,
i.e.\  $\widetilde F(\xi ):=\co F(\xi )$ for each $\xi \in \Xo$.
It is an easy exercise to verify that if $F$ is compact-valued, locally
Lipschitz, and locally bounded, then $\widetilde F$ also has these properties.

Clearly, if $\So$ is strongly invariant under $\widetilde F$ then it is also 
strongly invariant under $F$, because every solution of $\dot x\in  F(x)$ must also
be a solution of $\dot x\in  \widetilde F(x)$.
Conversely, suppose that $\So$ is strongly invariant under $F$, and consider
any solution $x:[0,T]\rightarrow \Xo$ of $\dot x\in  \widetilde F(x)$ which has the property that
$x(0)\in  \So$.
The Filippov-Wa\v{z}ewski Relaxation Theorem 
provides a sequence of solutions $x_k$, $k=1,2,\ldots $, of $\dot x\in  F(x)$
on the interval $[0,T]$,
with the property that $x_k(t)\rightarrow x(t)$ uniformly on $t\in [0,T]$ and also
$x_k(0)=x(0)\in  \So$ for all $k$.
Since $\So$ is strongly invariant under $F$, it follows that
$x_k(t)\in  \So$ for all $k$ and $t\in [0,T]$, and taking the limit as $k\rightarrow \infty $
this implies that also $x(t)\in  \So$ for all $t$.
In summary, invariance under $F$ or $\widetilde F$ are equivalent, for closed sets.

Let $N$ be a compact subset of $\Xo$ which contains $M$ in its interior
$\inter N$
and pick any smooth function $\varphi:\R^n\rightarrow \R_{\geq 0}$ with support equal to $N$
(that is, $\varphi(\xi )\equiv 0$ if $x\not\in  \inter N$ and $\varphi(\xi )>0$ on $\inter N$) and
such that $\varphi(\xi )\equiv 1$ on the set $M$.
Now consider the new differential inclusion defined on all of $\R^n$
given by $\hatF(\xi ):=\varphi(\xi )\widetilde F(\xi )$
if $\xi \in  N$ and equal to $\{0\}$ outside $N$.
Since $\widetilde F$ is locally Lipschitz and locally bounded, it follows by a standard
argument that $\hatF$ has these same properties.
Moreover, $\hatF$ is globally bounded and it is also convex-valued and
compact-valued (see e.g.~\cite{isw}).
Thus $\hatF$ is nice, as required.
Note that Property~(\ref{F-included-hatF-on-M})
holds, because $F(\xi )\subseteq \widetilde F(\xi )$ and $\varphi\equiv 1$ on $M$.

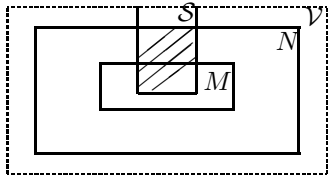
\begin{figure}[h,t]
\begin{center}
\setlength{\unitlength}{1200sp}%
\begin{picture}(6624,3506)(1789,-5173)
\thicklines
\put(2401,-4711){\framebox(5400,2550){}}
\put(3751,-3811){\framebox(2700,900){}}
\thinlines
\put(4501,-2461){\line( 5, 4){375}}
\put(4501,-2761){\line( 5, 4){750}}
\put(4501,-3061){\line( 5, 4){1125}}
\put(4501,-3361){\line( 5, 4){1125}}
\put(4801,-3436){\line( 4, 3){900}}
\put(5251,-3436){\line( 6, 5){450}}
\put(1801,-5161){\dashbox{60}(6600,3450){}} %{\framebox(6600,3450){}}
\thicklines
\put(4501,-1711){\line( 0,-1){1800}}
\put(4501,-3511){\line( 1, 0){1200}}
\put(5701,-3511){\line( 0, 1){1800}}
\put(5326,-2011){$\So$}
\put(7951,-2086){$\Xo$}
\put(7351,-2611){$N$}
\put(5850,-3436){$M$}
\end{picture}
\caption{Shaded area is set $\Sop$}
\label{M-N-sets}
\end{center}
\end{figure}

Let $\Sop:=\So\bigcap  N$
(cf.~Figure~\ref{M-N-sets});
this is a closed subset of $\R^n$ because the compact set
$N$ has a strictly positive distance to the complement of $\Xo$.
Property~(\ref{S-vs-S'})
holds as well, because $M\subseteq  N$.
Now pick any $\xi \in \Sop$.
There are two cases to consider:
$\xi $ is in the boundary of $N$ or in the interior of $N$.
If $\xi \in \partial N$, then $\hatF(\xi )=\{0\}$ because $\varphi(\xi )=0$.
If instead $\xi $ belongs to the interior of $N$, there is some open
subset $V\subseteq  N$ such that $\xi \in  V$.  Therefore
any sequence $\xi _i \rightarrow  \xi $ with all $\xi _i\in \So$ has, without loss of generality,
$\xi _i\in  V\bigcap  \So \subseteq  N\bigcap  \So = \Sop$, so also $\xi _i \rightarrow  \xi $ in $\Sop$; this
proves that $\tanSx\subseteq  \tanSpx$, 
and the reverse inclusion is true because $\Sop\subseteq \So$.
Hence Property~(\ref{possible-tangents}) has been established.
Regarding Property~(\ref{possible-tangents-M}), this follows from the
discussion in the previous paragraph, since $M$ is included in the interior of
$N$.

In order to prove the last property in the theorem,
we start by remarking that
if $x:[0,T]\rightarrow \R^n$ is a solution of $\dot x\in \hatF(x)$ with the property that
$x(t)$ belongs to the interior of $N$ for all $t$ (equivalently, $\varphi(x(t))\not= 0$
for all $t$), then there is a reparametrization of time such that $x$ is a
solution of $\dot x\in \widetilde F(x)$.  In precise terms: there is an interval $[0,R]$,
an absolutely continuous function $\alpha :[0,\infty )\rightarrow [0,\infty )$ such that $\alpha (0)=0$
and $\alpha (R)=T$, and a solution $z:[0,R]\rightarrow \R^n$ of $\dot z\in \widetilde F(z)$ such that
$z(r)=x(\alpha (r))$ for all $r\in [0,R]$.
To see this, it is enough (chain rule, remembering that
$\hatF(\xi )=\varphi(\xi )\widetilde F(x)$) for $\alpha $ to solve the initial value
problem $d\alpha /dr = \beta (\alpha (r))$, $\alpha (0)=0$, where $\beta (t)=1/\varphi(x(t))$ for $t\leq T$
and $\beta (t)\equiv \beta (T)$ for $t>T$.
The function $\varphi(x(t))$ is absolutely continuous, and is bounded away from
zero for all $t\leq T$ (because the solution $x$ lies in a compact subset of the
interior of the support of $\varphi$), so $\beta $ is locally Lipschitz and a (unique)
solution exists.  Since $\beta $ is globally bounded, the solution has no finite
escape times.  In addition, since the vector field is everywhere positive,
$\alpha (r)\rightarrow \infty $ as $s\rightarrow \infty $, so there is some $R$ such that $\alpha (R)=T$.

Now suppose that $\So$ is invariant under $F$.
As remarked, then $\So$ is invariant under its
convexification $\widetilde F$.
Suppose that $x:[0,T]\rightarrow \R^n$ is a solution of $\dot x=\hatF(x)$ such that
$x(0)\in \Sop$ and $x(t)$ is in the interior of $N$ for all $t$.
We find a solution $z$ of $\dot z\in \widetilde F(z)$ such that $z(r)=x(\alpha (r))$ for all
$r\in [0,R]$ and $z(0)=x(0)\in \Sop\subseteq  \So$ as earlier.
Invariance of $\So$ under $\widetilde F$ gives that $z(r)$, and hence $x(t)$, remains
in $\So$.
Since $\Sop=\So\bigcap  N$, we conclude that $x(t)\in \Sop$ for all $t\in [0,T]$.

Next, we use some ideas from the proof of Theorem 4.3.8 in~\cite{clsw}.
Pick any $\xi _0\in \Sop$, and any $v\in \hatF(\xi _0)$.
Define the mapping $f:\R^n\rightarrow \R^n$ by the following rule:
for each $\xi \in \R^n$, $f(\xi )$ is the unique closest point to $v$ in $\hatF(\xi )$.
As in the above citation, this map is continuous.
We claim that, for each $\xi \in \Sop$ there is some $\delta >0$ and a solution of
$\dot x=f(x)$ such that $x(0)=\xi $ and $x(t)\in \Sop$ for all $t\in [0,\delta ]$.
(Note that, in particular, this $x$ solves $\dot x\in \hatF(x)$.)
If $\xi $ is on the boundary of $N$, then $\hatF(\xi )=\{0\}$ implies that
$f(\xi )=0$, and hence $x(t)\equiv \xi $ is such a solution.
If instead $\xi $ belongs to the interior of $N$ then the previous remarks
shows that $x(t)\in \Sop$ for all $t\in [0,\delta ]$, where we pick a smaller $\delta $ if
needed in order to insure that $x(t)$ remains in the interior of $N$.
We conclude from the claim that the closed set $\Sop$ is locally-in-time
invariant with respect to the differential inclusion $\{f(x)\}$, which
satisfies the ``standing hypotheses'' in Chapter 4 of~\cite{clsw}.
This inclusion is hence also ``weakly invariant'' as follows from
Exercise~4.2.1 in that textbook.
This in turn implies, by Theorem~4.2.10 there, that
$\ip{f(\xi )}{\zeta }\leq 0$ for all $\xi \in \Sop$ and all $\zeta $ in the proximal normal
set $N_\xi \Sop$ defined in that reference (we are using a different notation).
Applied in particular at the point $\xi _0$ (so that $f(\xi _0)=v$), we conclude
that  $\ip{v}{\zeta }\leq 0$ for all $\zeta \in  N_{\xi _0}\Sop$.
Since $v$ was an arbitrary element of $\hatF(\xi _0)$, it follows that the upper
Hamiltonian condition in part (d) of Theorem 4.3.8 in~\cite{clsw} holds for
the map $\hatF$ at the point $\xi _0$.
Since $\xi _0$ was itself an arbitrary point in $\Sop$, the condition holds on
all of $\Sop$.
Therefore $\Sop$ is invariant for $\hatF$, as claimed.
\epr

\noindent{\bf Proof of Theorem~\protect{\ref{main-theorem-invariance}}}

We first prove that 2$\Rightarrow $1.
Suppose that $F(\xi )\subseteq  \tanSx$ for every $\xi \in  \So$, and
pick any solution $x:[0,T]\rightarrow \Xo$ of $\dot x\in F(x)$ with $x(0)\in \So$.

Since $x(\cdot )$ is continuous, there is some compact subset $M\subseteq \Xo$ such
that $x(t)\in  M$ for all $t\in [0,T]$.
We apply Lemma~\ref{main-reduction-invariance} to obtain $\hatF$ and $\Sop$.
By Property~(\ref{F-included-hatF-on-M}), it holds
that $x$ is also a solution of $\dot x\in \hatF(x)$, and
Property~(\ref{S-vs-S'}) gives that $x(0)$ belongs to the subset $\Sop$.
Taking convex hulls, $\widetilde F(x)\subseteq  \co\tanSx$ for every $x\in  \So$.
Since $\hatF$ is a scalar multiple of $\widetilde F$,
and $\co\tanSx$ is a cone (because $\tanSx$ is a cone),
it follows that $\hatF(\xi )\subseteq  \co\tanSx$ for every $\xi \in  \So$, and so also
for $\xi \in \Sop$.
By Property~(\ref{possible-tangents}),
$\hatF(\xi )\subseteq  \co\tanSpx\quad\quad\forall\,\xi \in  \Sop$,
since either $\hatF(\xi )=0$ or $\tanSpx=\tanSx$ (and hence their convex hulls
coincide).

In summary, Property~(\ref{co-tan}) is valid for $\hatF$ in place of $F$ and
$\Sop$ in place of $\So$, and $\hatF$ is nice.
Thus we may apply Theorem 4.3.8 in \cite{clsw} to conclude that $\Sop$ is
strongly invariant under $\hatF$. 
Since $x(0)\in \Sop$, it follows that $x(t)\in \Sop$ for all $t\in [0,T]$, and
therefore also $x(t)\in \So$ for all $t\in [0,T]$, as wanted.

We now prove that 1$\Rightarrow $2.
Suppose that $\So$ is strongly invariant under $F$, and pick any $\xi _0\in \So$.
We apply Lemma~\ref{main-reduction-invariance}, with $M=\{\xi _0\}$, to obtain
$\hatF$ and $\Sop$.
Note that $M\bigcap  \So=\{\xi _0\}$, so $\xi _0\in \Sop$.
Moreover, $\Sop$ is strongly invariant
under $\hatF$.
Since $\Sop$ is closed and $\hatF$ is nice, Theorem 4.3.8 in \cite{clsw}
gives that $\hatF(\xi )\subseteq \tanSpx$  for all $\xi \in \Sop$, and in particular for
$\xi =\xi _0$.
By Property~(\ref{possible-tangents}), either $\hatF(\xi _0)=\{0\}$
or $\tanSpx=\tanSx$, so we have that 
$\hatF(\xi )\subseteq \tanSx$ for $\xi =\xi _0$.
Moreover, 
Property~(\ref{F-included-hatF-on-M}) gives that $F(\xi ) \subseteq  \hatF(\xi )$
for $\xi =\xi _0$.
Since $\xi _0$ was an arbitrary element of $\So$, the proof is complete.
\qed

\end{document}